\documentclass[twocolumn]{aastex631}
\usepackage{mathtools}
\usepackage{amsmath}
\usepackage{natbib}
\usepackage{graphicx}
\usepackage{appendix}
\usepackage{amssymb}
\usepackage{float}
\usepackage{multirow}
\usepackage{booktabs}
\usepackage{colortbl}
\usepackage{makecell}
\graphicspath{{}}
\usepackage[x11names]{xcolor}
\usepackage{hyperref}
\hypersetup{colorlinks=true, citecolor=blue, linkcolor=cyan, urlcolor=magenta}
\usepackage[T1]{fontenc}
\usepackage{bm}
\usepackage{epstopdf} 
\usepackage{lipsum}
\usepackage{afterpage}

\newcommand{\rproj}{\ensuremath{R_{\rm proj}}}
\newcommand{\delv}{\ensuremath{\Delta V}}

\shorttitle{Environmental status of RQEs}
\shortauthors{D. K. Deo}

\begin{document}

\title{Are Recently Quenched Ellipticals Truly Isolated Centrals? \\ A Multi-scale Environmental Reassessment}

\author{Deepak K. Deo}
\affiliation{Department of Physics and Astronomy, University of Missouri--Kansas City, Kansas City, MO 64110, USA}
\email{dd9wn@umsystem.edu}

\begin{abstract}
Recently Quenched Ellipticals (RQEs) provide a valuable test case for disentangling intrinsic and environmental quenching, particularly because they are commonly classified as isolated central galaxies in low-mass halos. However, central/satellite assignments and isolation labels can vary across group catalogs, and such misclassifications can strongly bias physical interpretations. We present a uniform, physically motivated reassessment of the environments of 155 RQEs previously identified as centrals in an SDSS-based group catalog. We construct value-added neighbor catalogs via KD-tree searches and apply consistent thresholds in projected separation and line-of-sight velocity to (i) verify centrality, (ii) quantify isolation using a mass-ratio-based companion criterion, and (iii) identify potential pseudo-centrals via proximity to massive clusters. We find that 132/155 (85.2\%) RQEs remain true centrals, while 23/155 (14.8\%) are better interpreted as misidentified centrals with a more massive neighbor within $R_{\rm proj}\leq0.8$~Mpc and $\Delta V\leq250$~km~s$^{-1}$. Among the true centrals, 110/132 (83.3\%) satisfy our isolation criterion, and only one system meets our pseudo-central definition, indicating that direct cluster-scale preprocessing is rare for RQE centrals. Using projected number density and surface stellar mass density, we show that misidentified and non-isolated systems occupy systematically denser regimes than isolated true centrals. These results imply that while most RQEs ($\sim$71\%) are consistent with predominantly internal quenching in genuinely isolated centrals, a non-negligible minority ($\sim$29\%) likely experienced environmentally influenced pathways at group scales.
\end{abstract}


\section{Introduction}

Central galaxies in low-mass halos are generally expected to sustain star formation by continuing to accrete cold gas \citep{2005MNRAS.363....2K, 2009A&A...498..407G}. Recently Quenched Ellipticals (RQEs), however, represent a puzzling exception to this picture. These local ellipticals have recently quenched and, according to recent work, a subset may retain appreciable neutral hydrogen reservoirs, with inferred HI gas fractions reaching $\gtrsim 17\%$ \citep{deo2026investigatingquenchingrecentlyquenched}. The coexistence of suppressed star formation with substantial HI suggests that quenching in these systems may involve mechanisms beyond simple gas depletion, such as processes that heat, stabilize, or otherwise prevent the cold gas from efficiently collapsing into star-forming regions. Interpreting these possibilities, however, often requires a reliable assessment of environment and central/satellite status, since the same observed quiescence can imply very different physical origins depending on whether a galaxy is truly an isolated central or instead a satellite or misidentified system.

More generally, star formation quenching can occur through mechanisms that remove the cold-gas reservoir, heat it so that it cannot cool efficiently, or shut off the supply of fresh gas, thereby suppressing further star formation. These mechanisms are commonly discussed in terms of two broad channels: intrinsic processes tied to a galaxy's internal properties (often correlated with stellar mass), and environmental processes driven by interactions with surrounding structures and neighboring galaxies \citep{2010ApJ...721..193P}. Intrinsic (``mass'') quenching includes AGN feedback, stellar feedback, and structural/morphological stabilization \citep{1998A&A...331L...1S, 2005Natur.433..604D, 2012ARA&A..50..455F, 2018arXiv180909427F, 2012MNRAS.421.3522H, 2019MNRAS.487.3740W, 0004-637X-707-1-250}. Environmental quenching includes processes such as ram-pressure stripping, tidal interactions and mergers, high-speed harassment, and strangulation \citep{1972ApJ...176....1G, 1972ApJ...178..623T, 2008ApJS..175..356H, 1996Natur.379..613M, 1980ApJ...237..692L, 2015Natur.521..192P}. Because these channels can operate in overlapping regimes, robust inference about the dominant pathway for any given galaxy population depends sensitively on accurate environmental and central/satellite classification.

RQEs were identified by \citet{2014MNRAS.442..533M} (hereafter M14) as a population of local ($z \leq 0.08$) galaxies with elliptical morphology and very young light-weighted stellar ages (Age$_{\rm lw} \leq 3$~Gyr), indicative of substantial recent star formation followed by abrupt cessation. Their young stellar populations place them in the blue region of color--mass and color--color ($u-r$ versus $r-z$) diagrams, even though their morphologies are consistent with early-type systems. The elliptical morphology of RQEs has been interpreted as suggestive of a major-merger origin \citep{1972ApJ...178..623T, 1977egsp.conf..401T}, motivating the view that at least some systems may represent recently formed, rapidly quenched ellipticals. In the M14 sample, RQEs are commonly associated with low-mass halos, implying that quenching must have acted efficiently despite the expectation of continued cold-gas accretion in such environments.

Environmental interpretation for RQEs depends on reliable central--satellite classification. Using the halo-based group catalog of \citet{2007ApJ...671..153Y} (hereafter Y07), constructed from SDSS DR4 \citep{2006ApJS..162...38A}, M14 report that $\sim 90\%$ of RQEs are classified as centrals, residing in small groups with median halo masses $M_{\rm halo} \simeq 10^{12.2}\,M_\odot$. Moreover, Y07 classifies a large fraction of these central RQEs as ``isolated centrals,'' i.e., the only detected members of their groups within the SDSS DR4 limits. If correct, such classifications would favor intrinsic quenching mechanisms for most RQEs. However, independent group catalogs yield non-negligible discrepancies in the inferred environmental status of the same galaxies, introducing uncertainty in the physical interpretation.

For example, \citet{2017MNRAS.470.2982L} applied a modified friends-of-friends approach with variable linking lengths to SDSS DR13 and classified $\sim 86\%$ of Y07 central RQEs as centrals, with $\sim 65\%$ of those labeled isolated. Similarly, \citet{2020arXiv200712200T, 2021ApJ...923..154T} used a redshift-space group finder on SDSS DR7 and identified $\sim 94\%$ of Y07 central RQEs as centrals, with $\sim 76\%$ designated as isolated. Although these catalogs adopt broadly consistent concepts of centrality and isolation, differences in group-finding methodology, survey depth, and completeness can change group assignments and can lead to ``pseudo-centrals''---systems labeled as centrals in low-mass halos but actually associated with (or infalling into) more massive structures. Therefore, before using RQEs to discriminate between intrinsic and environmental quenching pathways, it is essential to verify whether they are truly isolated centrals or instead misidentified systems.

These considerations motivate a unified reassessment of RQE environments using consistent and physically motivated criteria. In this study, we develop and apply a uniform framework to re-evaluate the environmental context of the full sample of 155 Y07 central RQEs from M14. Our goals are to (i) verify centrality and identify potential misidentified centrals using thresholds based on projected separation, line-of-sight velocity difference, and stellar mass (and where available, halo mass) of neighboring galaxies; (ii) quantify environmental influence using multi-scale diagnostics, including projected local number density and surface stellar mass density; and (iii) connect these environmental regimes to HI constraints to test whether environmental effects can help explain quenching in gas-rich RQEs. By improving the reliability of the central/isolation classification and placing RQEs within a consistent multi-scale environmental framework, we aim to clarify when RQEs are best interpreted as intrinsically quenched centrals versus systems whose recent quenching may have been environmentally influenced.

This paper is organized as follows. Section~\ref{data2} describes the data sets and sample selection. Section~\ref{method2} presents our methodology for centrality verification and environmental diagnostics. Section~\ref{res2} reports the results, separating isolated true centrals, non-isolated true centrals, and misidentified centrals, and presenting their environmental metrics. Section~\ref{disc2} discusses implications for quenching pathways and for the interpretation of RQEs as transition systems. Section~\ref{conc2} summarizes our conclusions.

\section{Data and Sample}
\label{data2}

To investigate the environmental conditions that may contribute to the quenching of Recently Quenched Ellipticals (RQEs) in the local universe, we combine data from multiple observational sources. Our analysis begins with a mass-selected sample of RQEs identified from SDSS, and incorporates additional datasets to characterize their group membership, quantify their local galaxy environments, and assess proximity to massive galaxy clusters.

Specifically, we use SDSS-based galaxy group catalogs to evaluate the central or satellite status of each RQE, a spectroscopic galaxy catalog to identify neighboring galaxies and compute local environmental metrics, and a galaxy cluster catalog constructed from the DESI Legacy Surveys and WISE imaging to identify associations with massive halos. Below, we describe the construction of the RQE sample and each of these datasets in detail.

\subsection{Recently Quenched Elliptical Galaxy Sample}\label{ch2:rqe_sample}

Our primary sample consists of 172 RQEs identified by M14 from a complete, stellar mass-limited ($M_\star \geq 2 \times 10^{10}\,M_\odot$), and redshift-limited ($0.02 \lesssim z \lesssim 0.08$) sample of 63,454 galaxies drawn from the New York University Value-Added Galaxy Catalogue (NYU-VAGC; \citealt{2005AJ....129.2562B}), based on SDSS Data Release 4 \citep{2006ApJS..162...38A}.

M14 selected RQEs through a multi-step process: first isolating blue early-type galaxies using empirical color and concentration cuts, then employing visual classification to identify elliptical morphologies, and finally using emission-line diagnostics and light-weighted stellar ages (Age$_{\rm lw} \leq 3$~Gyr) to select recently quenched systems. Full details of the selection procedure are provided in M14 and \cite{deo2026investigatingquenchingrecentlyquenched}.

Of the 172 RQEs, 155 are classified as central galaxies in the halo-based group catalog of Y07, constructed from SDSS DR4. Y07 identifies centrals as the most massive or luminous galaxy within each group, determined through an iterative halo-mass assignment procedure. The remaining 17 RQEs are classified as satellites and are excluded from this study, as satellite RQEs are more likely to be quenched via well-established satellite-specific mechanisms such as ram-pressure stripping and tidal interactions \citep{1972ApJ...176....1G, 1972ApJ...178..623T, 2012ApJ...757....4P}.

The 155 central RQEs reside in low-mass dark matter halos with a median halo mass of $M_{\rm halo} \approx 1.5 \times 10^{12}\,M_\odot$, a regime where galaxies are typically expected to continue forming stars. Their unexpected quiescence, combined with evidence that a subset may retain appreciable neutral hydrogen reservoirs, makes them an intriguing population for investigating quenching mechanisms in low-density environments \citep{deo2026investigatingquenchingrecentlyquenched}. 

The final catalog incorporates value-added properties including HI content estimates, star formation rates, and NUV$-r$ colors from multi-wavelength analysis \citep{deo2026investigatingquenchingrecentlyquenched}, enabling exploration of connections between gas content, environment, and quenching.

\subsection{Galaxy Group Catalogs}
\label{sec:group_catalogs}
To investigate the discrepancies in environmental classifications of RQEs, we utilized three distinct SDSS-based galaxy group catalogs, each employing different methodologies and data releases:
\begin{itemize}
    \item \textbf{Yang et al. (2007) catalog:} This catalog uses a halo-based group finder algorithm applied to SDSS DR4 data. The method employs an iterative approach that identifies potential group centers, estimates their characteristic luminosity, and assigns tentative masses based on average mass-to-light ratios to determine group membership in redshift space \citep{2007ApJ...671..153Y}. The catalog includes 301,237 groups across a broad mass range. Central galaxies are identified as the most luminous members of their groups, with isolated centrals defined as the sole members of their respective groups. In this catalog, approximately 74\% of our RQEs are classified as isolated centrals.
    
    \item \textbf{Lim et al. (2017) catalog:} Based on SDSS DR13, this catalog employs a modified friends-of-friends algorithm with variable linking lengths to identify galaxy groups \citep{2017MNRAS.470.2982L}. When cross-matched with our sample, we found matches for only 86\% of our RQEs. All of these matched RQEs were classified as centrals, but only 65\% of the matched RQEs were designated as isolated centrals (56\% of our total sample).
    
    \item \textbf{Tinker et al. (2020, 2021) catalog:} This catalog applies a redshift-space group finder to SDSS DR7 data, optimized for clustering studies and incorporating color-dependent clustering to improve halo mass estimates \citep{2020arXiv200712200T, 2021ApJ...923..154T}. We found matches for all RQEs in our sample, with 94\% classified as centrals and 76\% of these centrals identified as isolated centrals.
\end{itemize}

These catalogs provide varying classifications for the same set of RQEs, highlighting inconsistencies in how centrality and isolation are determined. Table \ref{tab:group_class} summarizes the classification distributions across these catalogs.

\begin{deluxetable*}{lcccc}
\tablecaption{RQE Classification Across Different Group Catalogs \label{tab:group_class}}
\tablewidth{0pt}
\tablehead{
\colhead{Group Catalog} & \colhead{Match Rate (\%)} & \multicolumn{3}{c}{Of Matched RQEs} \\
\cline{3-5}
\colhead{} & \colhead{} & \colhead{Centrals (\%)} & \colhead{Isolated Centrals (\%)} & \colhead{Satellites (\%)}
}
\startdata
Yang et al. (2007)         & 100 & 100 & 74 & 0 \\
Lim et al. (2017)          & 86  & 100 & 65 & 0 \\
Tinker et al. (2020, 2021) & 100 & 94  & 76 & 6 \\
\enddata
\end{deluxetable*}

These discrepancies arise from differences in group-finding algorithms and varying data sensitivity across SDSS releases. The methodologies of Lim et al. and Tinker et al., which use variable linking lengths and redshift-space considerations, may detect group associations that Y07's fixed criteria miss. These inconsistencies in classification highlight the need for a consistent approach to accurately determine whether RQEs are truly isolated or environmentally influenced, which is essential for understanding their quenching mechanisms.

\subsection{Galaxy Catalog for Environment Characterization} \label{sec:galaxy_catalog}

To perform a detailed environmental analysis of RQEs, we utilized the SDSS DR13 galaxy catalog from \citet{2017MNRAS.470.2982L}, which provides positions, redshifts, and stellar mass estimates for galaxies in the SDSS footprint. This catalog served as the basis for identifying neighboring galaxies around each RQE within our defined thresholds for projected physical separation and velocity difference.

Using this catalog, we constructed a value-added catalog containing all RQE-neighbor pairs within a certain projected physical separation ($R_{\text{proj}}$) and a velocity difference ($\Delta V$). For each pair, we calculated various distance metrics and the mass ratio between the neighboring galaxy and its associated RQE to assess potential gravitational influences.

\subsection{Massive Cluster Catalog}\label{sec:cluster_catalog}
To identify potential pseudo-central RQEs, galaxies that appear to be centrals of small groups but may be influenced by or infalling into larger structures, we utilized the \citet{2024ApJS..272...39W} catalog of massive galaxy clusters. This catalog, based on the DESI Legacy Imaging Surveys and WISE, contains 1.58 million clusters with masses $M_{500} \geq 0.47 \times 10^{14} M_{\odot}$ and extends to higher redshifts than previous catalogs.

By cross-matching our RQEs with this cluster catalog, we identified those RQEs that have a massive cluster within their vicinity (defined by our $R_{\text{proj}}$ and $\Delta V$ thresholds). RQEs meeting these proximity criteria were flagged as potential pseudo-centrals for further analysis, enabling us to assess whether their quenching might be influenced by large-scale environmental effects despite their apparent central status in smaller groups.

Having assembled our primary sample of central RQEs along with the accompanying galaxy, group, and cluster catalogs, we next outline the methodology used to characterize their environments. In the following section, we describe how we construct the value-added catalogs used to identify neighboring galaxies, quantify environmental metrics, and evaluate each RQE’s centrality and isolation status. Together, these analyses provide the foundation for examining the role of environment in the quenching of recently quenched ellipticals.

\section{Environmental Analysis Methodology}\label{method2}
To investigate the environment of central RQEs, we developed a systematic framework to characterize their surroundings, centrality, and isolation using the datasets described in Section~\ref{data2}.
Our methodology consists of four interconnected components: (1) the creation of value-added catalogs through efficient neighbor identification using the KD-tree algorithm, (2) a comparative evaluation of isolation criteria from the literature and development of our unified criterion, (3) a robust assessment of centrality status and identification of potential pseudo-centrals, and (4) quantitative measurements of local environmental density.
Together, these steps enable a systematic reassessment of centrality and isolation designations for the RQE sample, addressing classification inconsistencies observed across different group catalogs and revealing the potential influence of the local environment on star formation quenching.

\subsection{Neighbor Identification Framework}
Characterizing the local environment of each RQE requires systematic identification of neighboring galaxies and nearby massive structures. Below, we describe the algorithm used for neighbor searches, the distance metric employed to rank neighbors, the procedure for removing spurious matches, and the resulting value-added catalogs.

\subsubsection{KD-Tree Algorithm Implementation}

Efficiently identifying neighboring galaxies and structures around each RQE requires an optimized computational approach. We implemented the KD-tree algorithm, a space-partitioning data structure that organizes points in a $k$-dimensional space \citep{bentley1975multidimensional}. This algorithm significantly reduces the computational complexity of nearest-neighbor searches from $O(n)$ to $O(\log n)$, where $n$ is the number of points in the dataset, enabling searches that would normally take hours to complete in seconds.

For our analysis, we applied the KD-tree algorithm to a three-dimensional representation of galaxy positions, using right ascension (RA), declination (DEC), and redshift ($z$) as coordinates. The algorithm recursively partitions the data space along each dimension, creating nested cells that optimize the search process. This implementation allowed us to rapidly identify all galaxies within specified projected physical separation ($R_{\text{proj}}$) and velocity difference ($\Delta V$) thresholds around each RQE.

We cross-matched our sample of 155 RQEs with the SDSS DR13-based galaxy catalog from \citet{2017MNRAS.470.2982L} to identify nearby galaxies within a projected physical separation of $R_{\text{proj}} \leq 3$ Mpc and a line-of-sight velocity difference of $\Delta V \leq 2000$ km s$^{-1}$. These thresholds were chosen to balance completeness and computational efficiency, while still encompassing a physically meaningful environment around each RQE where gravitational interactions and group-scale processes are likely to occur. Similar separation criteria have been employed in previous studies of galaxy environments \citep{2012ApJ...757...85G, 2015ApJ...809..146B, AF15}, particularly for assessing both local and large-scale environmental influences. 

We emphasize that these thresholds were optimized for identifying galaxy-scale neighbors in the field or small groups and differ slightly from the criteria used to identify proximity to massive clusters (see Section~\ref{sec:cluster_proximity}), where larger spatial and velocity separations are appropriate due to the extended influence of cluster-scale potentials.

To calculate the projected physical separation and velocity difference between galaxies, we used the following equations:

\begin{equation}
    R_{\text{proj}} = \theta \times D_A(z_{\text{RQE}})
\end{equation}

where $\theta$ is the angular separation in radians, and $D_A(z_{\text{RQE}})$ is the angular diameter distance at the redshift of the RQE. The velocity difference was calculated using the standard Doppler formula:

\begin{equation}
    \Delta V = c \times |z_{\text{RQE}} - z_{\text{neighbor}}|
\end{equation}

where $c$ is the speed of light, and $z_{\text{RQE}}$ and $z_{\text{neighbor}}$ are the redshifts of the RQE and neighboring galaxy or cluster, respectively. 

\subsubsection{Combined Distance Metric and Neighbor Ranking}

To quantify and rank the proximity of neighboring galaxies to each RQE, we developed a unified distance measure that incorporates both spatial and kinematic separation. This ``combined distance metric," $s_{\text{combined}}$, is defined as:

\begin{equation}
    s_{\text{combined}} = \sqrt{R_{\text{proj}}^2 + s_{\parallel}^2}
\end{equation}

where $s_{\parallel}$ represents the line-of-sight separation derived from velocity differences:

\begin{equation}
    s_{\parallel} = \frac{\Delta V}{H_0}
\end{equation}

with $H_0 = 70$ km s$^{-1}$ Mpc$^{-1}$ as the Hubble constant. This metric provides a single, physically meaningful measure of neighbor proximity in distance units (Mpc), accounting for both projected physical separation and line-of-sight velocity differences. The combined metric is particularly valuable for assessing environmental influence, as it naturally balances spatial closeness with kinematic association, giving a more complete picture of potential galaxy interactions.

For each RQE, neighboring galaxies were ranked based on their $s_{\text{combined}}$ values, with lower values indicating closer proximity and potentially stronger environmental influence. This ranking system enabled us to identify the most significant neighbors for each RQE, even in complex environments with multiple galaxies.

\subsubsection{Identification and Removal of Pseudo-Neighbors}

After generating the initial neighbor catalog using proximity criteria in projected separation and velocity space, we applied an additional refinement step to exclude \textit{pseudo-neighbors}$-$objects that, due to small separations in both position and redshift, are likely duplicate entries or alternate detections of the same physical galaxy rather than true companions. These false positives, if not removed, could bias the environmental characterization of RQEs by artificially inflating the number of close neighbors.

We adopted a two-step approach combining quantitative cuts and visual confirmation:

\begin{enumerate}
    \item \textbf{Quantitative Filtering:} We empirically identified thresholds for proximity below which neighbor entries were likely to be the RQE itself. Specifically, neighbors with line-of-sight velocity differences $\Delta V$ less than four times the stellar velocity dispersion of the RQE, and projected separations smaller than twice the Petrosian radius ($R_{90,r}$) of the RQE, were flagged as potential pseudo-neighbors. These criteria were established through trial-and-error testing on a pilot subsample (\(\sim\)10\% of RQEs), where such close companions consistently failed to represent distinct galaxies.
    
    \item \textbf{Visual Inspection:} For all flagged cases and for sanity checking, all top-ranked neighbors (i.e., those with the smallest $s_{\text{combined}}$ values), we performed manual inspection of SDSS images and spectra for both the RQE and its neighbor. Neighbors that shared identical imaging footprints or spectroscopic observations with the RQE were confirmed as pseudo-neighbors and excluded. This visual cross-check served to validate our automated filtering and to capture edge cases missed by the initial cuts.
\end{enumerate}

Through this filtering process, we identified and removed 131 pseudo-neighbors from the initial catalog of 2,545 RQE-neighbor pairs, representing approximately 5.1\% of all entries. These pseudo-neighbors were associated with 131 out of the 155 RQEs in the sample, with each affected RQE having one such spurious match. By combining empirically motivated selection criteria with targeted visual inspection, we ensured that the final refined catalog of 2,414 true RQE-neighbor pairs retained only physically distinct and environmentally meaningful galaxies for further analysis.

\subsubsection{Value-Added Catalogs}

To facilitate a detailed environmental characterization of RQEs, we constructed two value-added catalogs (VACs) based on the neighbor identification framework described above:

\begin{enumerate}

    \item \textbf{RQE-gal catalog:} This catalog contains 2,414 valid entries comprising 155 central RQEs and their respective neighboring galaxies, identified within a projected separation of $R_{\text{proj}} = 3$ Mpc and a line-of-sight velocity difference of $\Delta V = 2000$ km s$^{-1}$. Each entry includes galaxy coordinates, redshifts, stellar masses, multiple separation metrics (e.g., $R_{\text{proj}}$, $\Delta V$, $s_{\text{combined}}$), and the stellar mass ratio between the neighbor and its associated RQE. This dataset enables the quantitative assessment of local galaxy environments and the gravitational influence of nearby companions.

    \item \textbf{RQE-cluster catalog:} For each central RQE, we identified the nearest massive galaxy cluster ($M_{\text{halo}} \geq 10^{14} \, M_{\odot}$) using the \citet{2024ApJS..272...39W} catalog, which includes 1.58 million clusters based on DESI Legacy Surveys and WISE imaging, spanning halo masses $M_{500} \geq 0.47 \times 10^{14} \, M_{\odot}$ up to redshift $z < 1.5$. Using the same KD-tree framework and separation metrics as for galaxy neighbors, we matched each RQE to its nearest massive cluster within $R_{\text{proj}} = 4$ Mpc and $\Delta V = 3000$ km s$^{-1}$. These relatively generous thresholds were adopted to ensure sensitivity to large-scale gravitational effects, given that clusters can exhibit virial radii of up to $\sim$2-3 Mpc and velocity dispersions exceeding 1000 km s$^{-1}$ \citep{1996astro.ph.11148B}. Each entry includes the cluster's position, redshift, halo mass, and its spatial and velocity separation from the corresponding RQE, allowing us to identify potential pseudo-RQE-centrals influenced by nearby massive halos.

\end{enumerate}

These two catalogs form the foundation of our subsequent environmental analyses, providing a comprehensive view of each RQE's surroundings across multiple physical scales$-$from galaxy-scale interactions to large-scale cluster environments. This dual approach is essential for re-evaluating centrality and isolation classifications and for uncovering plausible pathways for quenching in these recently quenched systems.

\subsection{Evaluation and Definition of Galaxy Isolation Criteria}\label{sec:isolation_method}

In this subsection, we (i) summarize common isolation definitions used in the literature, (ii) apply a representative subset of those criteria to a pilot set of RQEs to quantify classification sensitivity, and (iii) define a unified, physically motivated isolation criterion tailored to our low-redshift, massive RQE sample.

\subsubsection{Review of Isolation Definitions in the Literature}

Accurately assessing whether a galaxy is isolated is essential for understanding its evolutionary path, particularly in studies of star formation quenching where environmental mechanisms may play a pivotal role. However, the definition of an ``isolated galaxy" varies widely in the literature, depending on the scientific context and data availability. Most definitions use combinations of projected physical separation ($R_{\text{proj}}$), line-of-sight velocity difference ($\Delta V$), and brightness or stellar mass contrasts to quantify the presence and potential influence of nearby galaxies.

One of the foundational works in this area is the Catalog of Isolated Galaxies (CIG) by \citet{1973AISAO...8....3K, 1980SvA....24..665K}, which defines an isolated galaxy based on angular diameter and projected angular distance. According to this catalog, a galaxy with angular diameter \( d \) is deemed isolated if all neighboring galaxies with angular diameters \( d_1 \) satisfying \( \frac{1}{4} d \leq d_1 \leq 4 d \) are at least \( 20 d_1 \) away in projected angular distance.

\cite{1995MNRAS.275...56N} included mass considerations by defining an isolated spiral galaxy as one having no galaxy more massive than itself within a radius of 1 Mpc. \cite{2001AJ....121..808C} analyzed isolated early-type galaxies (ETGs) and defined isolation as having no neighboring galaxies within a projected separation of \( 1 \, h^{-1}_{100} \) Mpc and a velocity difference of 1000 km s\(^{-1}\).

Further refinements incorporated magnitude and stellar mass differences. \cite{2003ApJ...598..260P} stipulated that an isolated galaxy must have no other galaxies within a magnitude difference \( \Delta M_B = 2 \), a projected separation of \( 500 \, h^{-1} \) kpc (with \( h = 0.7 \)), and a velocity separation of 1000 km s\(^{-1}\). \citet{2004MNRAS.354..851R} defined isolation for ETGs as having no neighbors within \( R_{\text{proj}} = 0.67 \) Mpc and \( \Delta V = 700 \) km s\(^{-1}\), and being at least 2.0 magnitudes brighter in the \( B \)-band than any nearby galaxy.

With the advent of large-scale surveys like the Sloan Digital Sky Survey (SDSS), isolation metrics have been refined using high-resolution data. \cite{2012MNRAS.424.2574W, 2014MNRAS.442.1363W} examined the abundance of satellite galaxies around bright isolated galaxies, defining an isolated galaxy as one with no brighter galaxies within \( R_{\text{proj}} = 1 \) Mpc and \( \Delta V = 1000 \) km s\(^{-1}\), and being at least one magnitude brighter than any other galaxy within \( R_{\text{proj}} = 0.5 \) Mpc and the same \( \Delta V \).

\cite{AF15} utilized SDSS Data Release 10 (DR10) to identify isolated galaxies, pairs, and triplets in the local universe (\( z \leq 0.08 \)), defining an isolated galaxy as one with no neighbors within \( R_{\text{proj}} = 1 \) Mpc and \( \Delta V = 500 \) km s\(^{-1}\). 
\citet{2015ApJ...809..146B} studied galaxies in low-density environments and defined isolation for high-mass systems as follows: a high-mass galaxy (\( M_{\star} \geq 10^{9.5} \, M_{\odot} \)) is considered isolated if it has no massive neighbors (defined as those with stellar masses at least 0.5 dex higher than the central galaxy) within a projected radius of 1.5 Mpc and a velocity difference of 1000 km s\(^{-1}\). Additionally, for galaxies with at least five neighbors, fifth nearest neighbor surface density ($\Sigma_5$) must be less than 1 Mpc$^{-2}$ to be considered isolated.

These studies illustrate the diverse methodologies employed to determine galactic isolation, each contributing to a broader understanding of galaxy evolution in different environments. A concise summary of these key selection criteria is provided in Table \ref{tab:literature_isolated_galaxies} to facilitate comparison and contextualization.

\begin{deluxetable*}{ll}
\tablecaption{Summary of Literature Definitions for Isolated Galaxies \label{tab:literature_isolated_galaxies}}
\tablewidth{0pt}
\renewcommand{\arraystretch}{1.5}
\tablehead{
\colhead{Study} & \colhead{Isolation Definition}
}
\startdata
\citet{AF15} & \parbox[t]{12cm}{No neighbors within $R_{\rm proj} = 1$~Mpc and $\Delta V = 500$~km~s$^{-1}$.\vspace{6pt}} \\
\hline
\citet{2015ApJ...809..146B} & \parbox[t]{12cm}{Massive galaxies ($M_\star \geq 10^{9.5}\,M_\odot$) are isolated if no galaxies within $R_{\rm proj} = 1.5$~Mpc and $\Delta V = 1000$~km~s$^{-1}$ are $\geq$0.5~dex more massive; for galaxies with 5+ neighbors, also requires $\Sigma_5 < 1$~Mpc$^{-2}$.\vspace{6pt}} \\
\hline
\citet{2012MNRAS.424.2574W, 2014MNRAS.442.1363W} & \parbox[t]{12cm}{No brighter galaxy within $R_{\rm proj} = 1$~Mpc and $\Delta V = 1000$~km~s$^{-1}$; target must also be the brightest by 1~mag in SDSS $r$-band within $R_{\rm proj} = 0.5$~Mpc and the same $\Delta V$.\vspace{6pt}} \\
\hline
\citet{2012ApJ...757...85G} & \parbox[t]{12cm}{Dwarf galaxies are considered isolated if no neighbor more massive than $2.5 \times 10^{10}\,M_\odot$ within a comoving distance of 1.5~Mpc and $\Delta V = 1000$~km~s$^{-1}$, within a search radius of 7~Mpc.\vspace{6pt}} \\
\hline
\citet{2004MNRAS.354..851R} & \parbox[t]{12cm}{No neighbors within $R_{\rm proj} = 0.67$~Mpc and $\Delta V = 700$~km~s$^{-1}$, and at least 2.0~mag brighter in $B$-band than any nearby galaxy.\vspace{6pt}} \\
\hline
\citet{2003ApJ...598..260P} & \parbox[t]{12cm}{No neighbors within a magnitude difference \(\Delta M_B = 2\), projected separation \(R_{\text{proj}} = 714 \text{ kpc}\) (adjusted from \(500 h^{-1} \text{ kpc}\) by dividing by \(h=0.7\)), and velocity separation \(\Delta V = 1000 \text{ km s$^{-1}$}\).\vspace{6pt}} \\
\hline
\citet{2001AJ....121..808C} & \parbox[t]{12cm}{No neighbors within \(R_{\text{proj}} = 1.43 \text{ Mpc}\) (adjusted from \(1h^{-1}_{100} \text{ Mpc}\) by dividing by \(h=0.7\)) and \(\Delta V = 1000 \text{ km s$^{-1}$}\).\vspace{6pt}} \\
\hline
\citet{1995MNRAS.275...56N} & \parbox[t]{12cm}{No neighbor more massive than itself within a radius of 1~Mpc.\vspace{6pt}} \\
\hline
\citet{1973AISAO...8....3K, 1980SvA....24..665K} & \parbox[t]{12cm}{A galaxy of angular diameter \(d\) is isolated if all neighboring galaxies with angular diameter \(d_1\) within \( \frac{1}{4}d \leq d_1 \leq 4d \) are at least \(20d_1\) away in projected angular distance.\vspace{6pt}} \\
\enddata
\end{deluxetable*}

\subsubsection{Application of Existing Isolation Criteria to the RQE Pilot Sample} \label{sec:isolation_lit}

From the many definitions in the literature, we selected three isolation criteria that are especially relevant for our RQE sample$-$massive ($M_{\star} > 10^{10} \, M_\odot$) galaxies in the local universe ($z < 0.08$) observed within the SDSS footprint:

\begin{enumerate}
    \item \textbf{\citet{AF15} (AF15):} No neighbors within $R_{\text{proj}} = 1$ Mpc and $\Delta V = 500$ km s$^{-1}$.
    \item \textbf{\citet{2015ApJ...809..146B} (B15):} No neighbors more massive by $\geq 0.5$ dex within $R_{\text{proj}} = 1.5$ Mpc and $\Delta V = 1000$ km s$^{-1}$. Also requires $\Sigma_5 < 1$ Mpc$^{-2}$ for galaxies with $\geq 5$ neighbors.
    \item \textbf{\citet{2014MNRAS.442.1363W} (W14):} No brighter galaxy within $R_{\text{proj}} = 1$ Mpc and $\Delta V = 1000$ km s$^{-1}$, and must be at least 1 mag brighter than any galaxy within $R_{\text{proj}} = 0.5$ Mpc and same $\Delta V$.
\end{enumerate}

We applied these three isolation criteria to a pilot sample of 20 randomly selected RQEs to assess the consistency of classification across methods. Table~\ref{tab:isol_stat} summarizes the isolation status of each RQE according to the criteria of AF15, B15, and W14, along with their designation in the Y07 group catalog.

\begin{deluxetable*}{lcccc}
\tablecaption{Isolation Status of 20 RQEs Across Different Classification Methods \label{tab:isol_stat}}
\tablewidth{0pt}
\tablehead{
\colhead{RQE ID} & \colhead{AF15} & \colhead{B15} & \colhead{W14} & \colhead{Y07}
}
\startdata
\textbf{nyu641550}  & \textbf{isolated}     & \textbf{isolated}     & \textbf{isolated}     & \textbf{isolated} \\
\textbf{nyu383407}  & \textbf{isolated}     & \textbf{isolated}     & \textbf{isolated}     & \textbf{isolated} \\
nyu913614  & non-isolated & non-isolated & non-isolated & isolated \\
nyu657018  & non-isolated & isolated     & isolated     & isolated \\
\textbf{nyu143489}  & \textbf{isolated}     & \textbf{isolated}     & \textbf{isolated}     & \textbf{isolated} \\
\textbf{nyu194598}  & \textbf{non-isolated} & \textbf{non-isolated} & \textbf{non-isolated} & \textbf{non-isolated} \\
nyu492817  & isolated     & non-isolated & isolated     & isolated \\
nyu394786  & isolated     & isolated     & isolated     & non-isolated \\
nyu288709  & non-isolated & non-isolated & non-isolated & isolated \\
\textbf{nyu154837}  & \textbf{isolated}     & \textbf{isolated}     & \textbf{isolated}     & \textbf{isolated} \\
nyu561726  & non-isolated & isolated     & non-isolated & isolated \\
nyu841903  & isolated     & isolated     & isolated     & non-isolated \\
nyu194768  & non-isolated & isolated     & isolated     & isolated \\
nyu374908  & isolated     & isolated     & isolated     & non-isolated \\
nyu928105  & isolated     & isolated     & non-isolated & isolated \\
nyu584761  & non-isolated & non-isolated & non-isolated & isolated \\
nyu307981  & isolated     & isolated     & non-isolated & isolated \\
nyu184761  & non-isolated & isolated     & isolated     & non-isolated \\
\textbf{nyu204862}  & \textbf{isolated}     & \textbf{isolated}     & \textbf{isolated}     & \textbf{isolated} \\
nyu509276  & non-isolated & isolated     & isolated     & isolated \\
\enddata
\tablecomments{Bold rows indicate RQEs where all four classification methods agree. AF15: \citet{AF15}; B15: \citet{2015ApJ...809..146B}; W14: \citet{2014MNRAS.442.1363W}; Y07: \citet{2007ApJ...671..153Y}.}
\end{deluxetable*}

The results reveal significant discrepancies. Only 60\% of the RQEs were consistently classified as either isolated or non-isolated across all three literature-based definitions. This consistency drops to just 30\% when compared against the Y07 classification. These findings underscore the sensitivity of isolation status to the specific thresholds and conditions adopted in different studies.

For example, an RQE classified as isolated by AF15 may fail the B15 criterion if it has a more massive neighbor located beyond the 1 Mpc threshold used by AF15 but within the broader 1.5 Mpc limit specified by B15. Similarly, the W14 criterion$-$which incorporates brightness differences$-$may designate an RQE as non-isolated if a brighter galaxy lies within its proximity cuts, even if mass-based or radial criteria are otherwise satisfied.

These discrepancies highlight how significantly isolation classifications depend on the specific parameters employed: the choice of $R_{\text{proj}}$ and $\Delta V$ thresholds, whether mass ratios or luminosity differences are considered, and the specific threshold values all substantially impact outcomes. With such variability in classification based on seemingly minor methodological differences, interpreting the true environmental influence on RQE evolution becomes challenging.

The substantial disagreement across established methods underscored the need for a unified, physically motivated criterion better aligned with the dynamical and gravitational processes most likely to influence the evolution of recently quenched ellipticals in low-density environments.

\subsubsection{Development of a Unified Isolation Criterion}

Drawing on insights from both the literature and our pilot sample results, we developed a unified isolation criterion tailored to determining whether a central RQE in a group or low-density environment is truly isolated$-$that is, free from significant gravitational influences that might affect its gas dynamics, star formation, and morphology.

We define a central RQE as \textbf{isolated} if, within $R_{\text{proj}} \leq 0.5$ Mpc and $\Delta V \leq 200$ km s$^{-1}$, there are no neighboring galaxies with stellar mass $M_{\text{neighbor}} \geq 0.1 M_{\text{central}}$.

This definition is motivated by three key considerations:

\begin{itemize}
    \item \textbf{Physical Separation} ($R_{\text{proj}} \leq 0.5$ Mpc): This radius represents the region where direct gravitational interactions most significantly influence galaxy properties. At approximately half the typical virial radius of small groups \citep{1996astro.ph.11148B}, this threshold focuses on the immediate environment where satellite galaxies can exert substantial gravitational effects on the central galaxy.
    
    \item \textbf{Dynamical Association} ($\Delta V \leq 200$ km s$^{-1}$): This velocity threshold aligns with the typical internal velocity dispersion of small galaxy groups \citep{1996astro.ph.11148B}. Galaxies with relative velocities below this threshold are likely dynamically associated and experience prolonged interaction periods, increasing their potential influence.
    
    \item \textbf{Mass Ratio} ($M_{\text{neighbor}} \geq 0.1 M_{\text{central}}$): This threshold identifies neighbors massive enough to exert significant gravitational influence through tidal interactions or minor mergers. The literature establishes that neighbors with mass ratios of 1:10 or greater can drive substantial evolutionary changes in central galaxies \citep{2008MNRAS.384..386C, 2009ApJ...699L.178N, 2021ApJ...912...45N}.
\end{itemize}

These criteria collectively identify ``influential neighbors"$-$galaxies capable of significantly affecting a RQE's evolution through direct gravitational interactions. Central RQEs lacking such influential neighbors within the specified thresholds are classified as truly isolated central systems.

\subsection{Assessment of Centrality Status}
A central goal of this work is to verify whether RQEs classified as centrals in group catalogs are in fact the most massive galaxies in their local halo and to identify systems whose apparent centrality may be misleading. In this subsection, we (i) apply a uniform, mass-based centrality check on group scales and (ii) flag ``pseudo-centrals,'' galaxies that satisfy the group-scale centrality criterion but lie close enough to a massive cluster that their evolution may be affected by cluster-scale environmental processes.

\subsubsection{Centrality Verification}

Accurately determining whether the central RQEs in our sample are truly the most massive galaxies within their respective groups is essential for understanding their evolutionary pathways. Although the Y07 group catalog classifies these RQEs as centrals, such designations can vary across catalogs due to differing methodologies. To mitigate this uncertainty, we implemented a systematic verification of centrality status using a physically motivated criterion.

We define an RQE as a \textbf{true central galaxy} if it is the most massive galaxy within a projected radius of $R_{\text{proj}} \leq 0.8$ Mpc and a line-of-sight velocity difference of $\Delta V \leq 250$ km s$^{-1}$. These thresholds are motivated by the following considerations:

\begin{itemize}
    \item \textbf{Physical Separation} ($R_{\text{proj}} \leq 0.8$ Mpc): Galaxy groups have characteristic radii ranging from 0.2 to 1.4 Mpc, with an average around 0.8 Mpc \citep{1996astro.ph.11148B}. Within this radius, member galaxies are expected to be gravitationally bound and share a common dark matter halo.

    \item \textbf{Dynamical Association} ($\Delta V \leq 250$ km s$^{-1}$): Small galaxy groups typically have velocity dispersions between 100 and 500 km s$^{-1}$, with a median around 250 km s$^{-1}$ \citep{1996astro.ph.11148B}. A velocity difference within this range suggests that galaxies are dynamically associated and not chance alignments.
\end{itemize}

Under this definition, any RQE with a more massive galaxy located within the specified spatial and velocity thresholds is flagged as a \textbf{potentially misidentified central}$-$that is, a galaxy originally classified as central in the Y07 catalog, but which may instead be a satellite of a more massive neighboring galaxy. This reassessment provides a more reliable interpretation of each RQE's evolutionary context, particularly in relation to environmental quenching mechanisms.

\subsubsection{Identification of Pseudo-Centrals}\label{sec:cluster_proximity}

Beyond verifying whether RQEs are true centrals within their immediate group-scale environment, we also identified galaxies we refer to as \textit{pseudo-centrals}. These systems appear to be centrals of small groups but may, in fact, be influenced by or actively infalling into larger cluster-sized halos. Identifying pseudo-centrals is crucial, as their proximity to massive clusters could subject them to environmental quenching processes like ram-pressure stripping, tidal interactions, and harassment, which operate on larger scales.

We classified an RQE as a pseudo-central if it satisfied our centrality criteria (i.e., the most massive galaxy within $R_{\text{proj}} \leq 0.8$ Mpc and $\Delta V \leq 250$ km s$^{-1}$), but was located within the sphere of influence of a nearby massive cluster. Specifically, an RQE was identified as a pseudo-central if there was a cluster-sized halo ($M_{\text{halo}} \geq 10^{14} , M_{\odot}$) within $R_{\text{proj}} \leq 2$ Mpc and $\Delta V \leq 2000$ km s$^{-1}$. These thresholds reflect typical cluster properties: galaxy clusters can have virial radii extending up to approximately 2 Mpc and velocity dispersions often exceeding 1000 km s$^{-1}$ \citep{1996astro.ph.11148B}, enabling their environmental influence to extend significantly beyond their immediate regions.

Through this classification, we aim to precisely differentiate truly isolated central RQEs from those potentially influenced by nearby massive structures. This distinction is critical for identifying galaxies whose quenching mechanisms may be driven primarily by large-scale environmental factors rather than internal processes or local group dynamics. A summary of the classification criteria for central, isolated central, and pseudo-central RQEs is provided in Table~\ref{tab:centrality_summary} for ease of reference.

\begin{deluxetable*}{ll}
\tablecaption{Summary of Environmental Classification Criteria for RQEs \label{tab:centrality_summary}}
\tablewidth{0pt}
\tablehead{
\colhead{Classification} & \colhead{Definition Criteria}
}
\startdata
Central RQE & Most massive galaxy within $R_{\mathrm{proj}} \leq 0.8$ Mpc and $\Delta V \leq 250$ km s$^{-1}$. \\
Isolated Central RQE & Central RQE with no influential neighbors ($M \geq 0.1\,M_{\mathrm{RQE}}$) within $R_{\mathrm{proj}} \leq 0.5$ Mpc, $\Delta V \leq 200$ km s$^{-1}$. \\
Pseudo-Central RQE & Central RQE within $R_{\mathrm{proj}} \leq 2$ Mpc and $\Delta V \leq 2000$ km s$^{-1}$ of a massive cluster ($M_{\mathrm{halo}} \geq 10^{14}\,M_\odot$). \\
\enddata
\end{deluxetable*}

\subsection{Quantifying Environmental Influence}\label{sec:envmetrics}

To understand the degree of isolation or environmental influence on Recently Quenched Ellipticals (RQEs), it is essential to quantify the surrounding environment in a systematic way. This quantification is crucial for investigating whether external factors such as nearby galaxies might impact the RQEs' gas content, star formation, and overall evolution. Two widely used approaches for environmental measurement in galaxy studies are the nearest neighbor approach \citep{2006MNRAS.373..469B, 2007ApJ...658..898P, 2008ApJ...674L..13C, 2011MNRAS.411.1869L} and the fixed aperture approach \citep{2005MNRAS.356.1155C, 2009ApJ...690.1883G, 2011MNRAS.411..929G}. Both methods offer distinct insights, each adapted to capture the density of galaxies around a target, either in compact groups or within broader fields. An overview of these two approaches and various adaptations can be found in \citet{2012MNRAS.419.2670M}.

The \textbf{nearest neighbor approach} estimates local density by identifying the closest neighboring galaxies to a target and using their distances to calculate a local density. A commonly employed metric is the projected distance to the \( n \)-th nearest neighbor, \( r_n \), which can be used to calculate the projected local density (or surface density) of galaxies, \(\sigma_n\), as:
\[
\sigma_n = \frac{n}{\pi r_n^2}
\]
where \( n \) is the number of neighbors within the projected distance \( r_n \).

The \textbf{fixed aperture approach}, on the other hand, involves placing a circular or spherical aperture of a fixed physical or angular size around the target galaxy and counting the number of galaxies within this aperture. The density, \(\sigma_n\), is then given by:
\[
\sigma_n = \frac{N}{\pi R^2}
\]
where \( N \) is the number of galaxies within the aperture of radius \( R \).

To best characterize the environment of RQEs$-$whether isolated or clustered$-$we use two specific parameters: \textbf{projected local number density} and \textbf{surface stellar mass density}. These metrics are tailored to reflect the low-density environments where RQEs are typically found, allowing for a nuanced assessment of how sparse or dense these regions are relative to isolated galaxy criteria.

\subsubsection{Projected Local Number Density}

The projected local number density is a measure of the number of neighboring galaxies around an RQE within a specific projected area. Historically, \citet{1980ApJ...236..351D} pioneered the study of galaxy morphology in dense environments, revealing a relationship between local density and galaxy type. He found that elliptical and S0 galaxies were more common in dense regions, while spirals dominated less dense environments. Dressler introduced a method that calculates projected local number density based on the distance to the 10th nearest neighbor, (\(R_{10}\)) as: 
\[
\frac{\text{number of neighbors} (10)}{\text{Area} (\pi R_{10}^{2})}
\]
This approach provided a robust measure of density in rich clusters and highlighted the morphological segregation of galaxies based on their local environments. However, Dressler's method is more suited to dense, clustered environments rather than the field or small group environments where RQEs are typically found \citep{1981ApJ...246L...5B}.

Given that RQEs often reside in small groups or low-density environments, we have adapted the fixed aperture approach for our analysis. We define the \textbf{projected local number density}, $\sigma_n$, as the number of galaxies within a circular area of radius \rproj, centered on the RQE, divided by the area of the circle:
\begin{equation}
   \sigma_n = \frac{N}{\pi R_{\text{proj}}^2}
\end{equation}
where $N$ is the number of neighboring galaxies within the projected radius \rproj  and within a line-of-sight velocity difference \delv  relative to the RQE. We have chosen \rproj and \delv to be consistent with the values used earlier in evaluating the centrality of RQEs (i.e., \rproj $\leq$ 0.8 Mpc and \delv $\leq$ 250 km s$^{-1}$), ensuring a coherent analysis framework.

\subsubsection{Surface Stellar Mass Density}

While the number density quantifies the abundance of neighboring galaxies, it does not account for their masses, which determine the gravitational influence they exert. To address this, we calculated the surface stellar mass density, $\Sigma_{\ast}$, defined as:

\begin{equation}
   \Sigma_{\ast} = \frac{\sum_{i} M_{\ast, i}}{\pi R_{\text{proj}}^2}
\end{equation}

where the sum is over all galaxies within a specified region ($R_{\text{proj}} = 0.8$ Mpc and $\Delta V \leq 250$ km s$^{-1}$), including the RQE itself, and $M_{\ast, i}$ is the stellar mass of the $i$-th galaxy. This metric accounts for the total stellar mass in the immediate environment, providing a mass-weighted perspective of the local density.

Studying the interplay between projected local number density and surface stellar mass density provides valuable insight into the nature of an RQE's environment and the possible external influences on its evolution. 
While a high number density ($\sigma_n$) and high surface stellar mass density ($\Sigma_\ast$) typically signal a dense, group or cluster environment$-$where gravitational interactions, tidal effects, or gas stripping processes are more likely to quench star formation$-$other combinations of these metrics reveal additional environmental complexities. For instance, a high $\Sigma_\ast$ accompanied by a relatively low $\sigma_n$ suggests that the local environment is sparsely populated but contains a few very massive galaxies. In such cases, the gravitational potential of nearby massive companions could still significantly perturb the RQE's gas content and star formation without requiring a high number of neighbors. Conversely, an environment with high $\sigma_n$ but low $\Sigma_\ast$ might indicate a crowding of low-mass galaxies, which may exert weaker individual influence but could still foster galaxy-galaxy interactions or harassment in aggregate. Finally, low values of both $\sigma_n$ and $\Sigma_\ast$ support the interpretation of a truly isolated system, where internal mechanisms are more likely to dominate the quenching process. Together, these two density metrics offer a complementary and physically motivated framework for probing the scale and nature of environmental effects acting on RQEs.

\section{Results}\label{res2}
We present the outcomes of our environmental reassessment in four steps. We first verify the group-scale centrality of the 155 Y07-classified central RQEs and identify misidentified centrals. We then quantify the isolation status of the true-central subset, test for proximity to massive clusters to flag pseudo-centrals, and finally characterize local environments using projected number and stellar-mass surface densities.

\subsection{Reassessment of RQE Environmental Classification}
In this subsection, we refine the environmental labels of the RQE sample by sequentially verifying centrality, quantifying isolation among true centrals, and testing for possible cluster-scale influence.

\subsubsection{Centrality Verification}

The Y07 group catalog classifies all 155 RQEs in our sample as central galaxies$-$that is, as the most massive galaxies within their respective groups. However, centrality assignments based solely on group-finding algorithms can be affected by projection effects, group membership uncertainties, and catalog-specific assumptions. Given the importance of accurately establishing centrality for environmental quenching studies, we re-evaluated the central status of each RQE using a physically motivated criterion.

We define a \textbf{true central} as an RQE that is the most massive galaxy within a projected radius of $R_{\text{proj}} \leq 0.8$ Mpc and a line-of-sight velocity difference of $\Delta V \leq 250$ km s$^{-1}$. If a more massive galaxy exists within this radius and velocity window, the RQE is flagged as a \textbf{potentially misidentified central}.

\begin{figure*}[ht!]
    \centering

    % First row: Two panels
    \begin{minipage}[t]{0.49\textwidth}
        \centering
        \includegraphics[width=\textwidth]{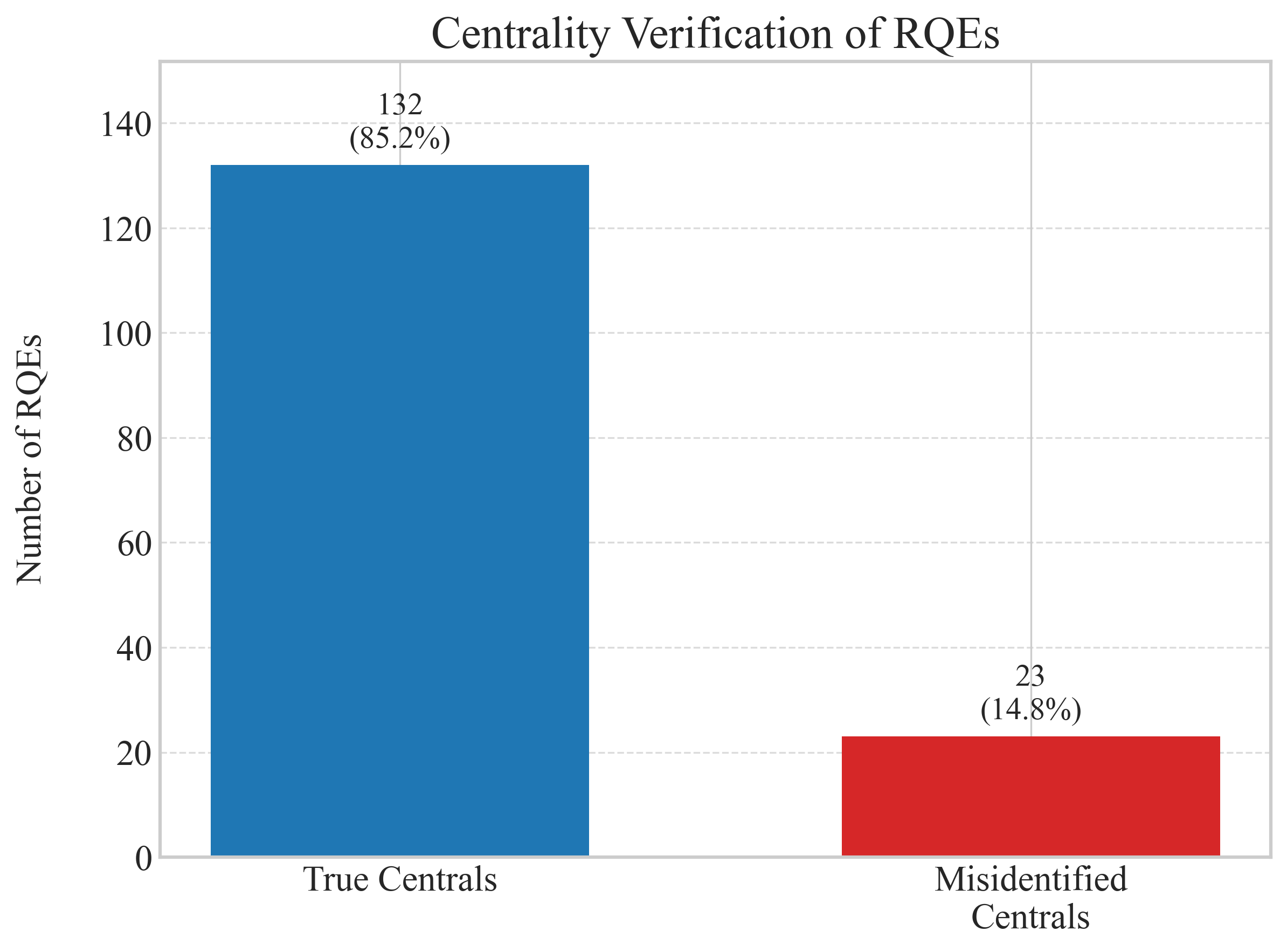}\\
        \vspace{2pt}
        {\small\textbf{(a)} Distribution of true and misidentified centrals in the RQE sample.}
    \end{minipage}
    \hfill
    \begin{minipage}[t]{0.49\textwidth}
        \centering
        \includegraphics[width=\textwidth]{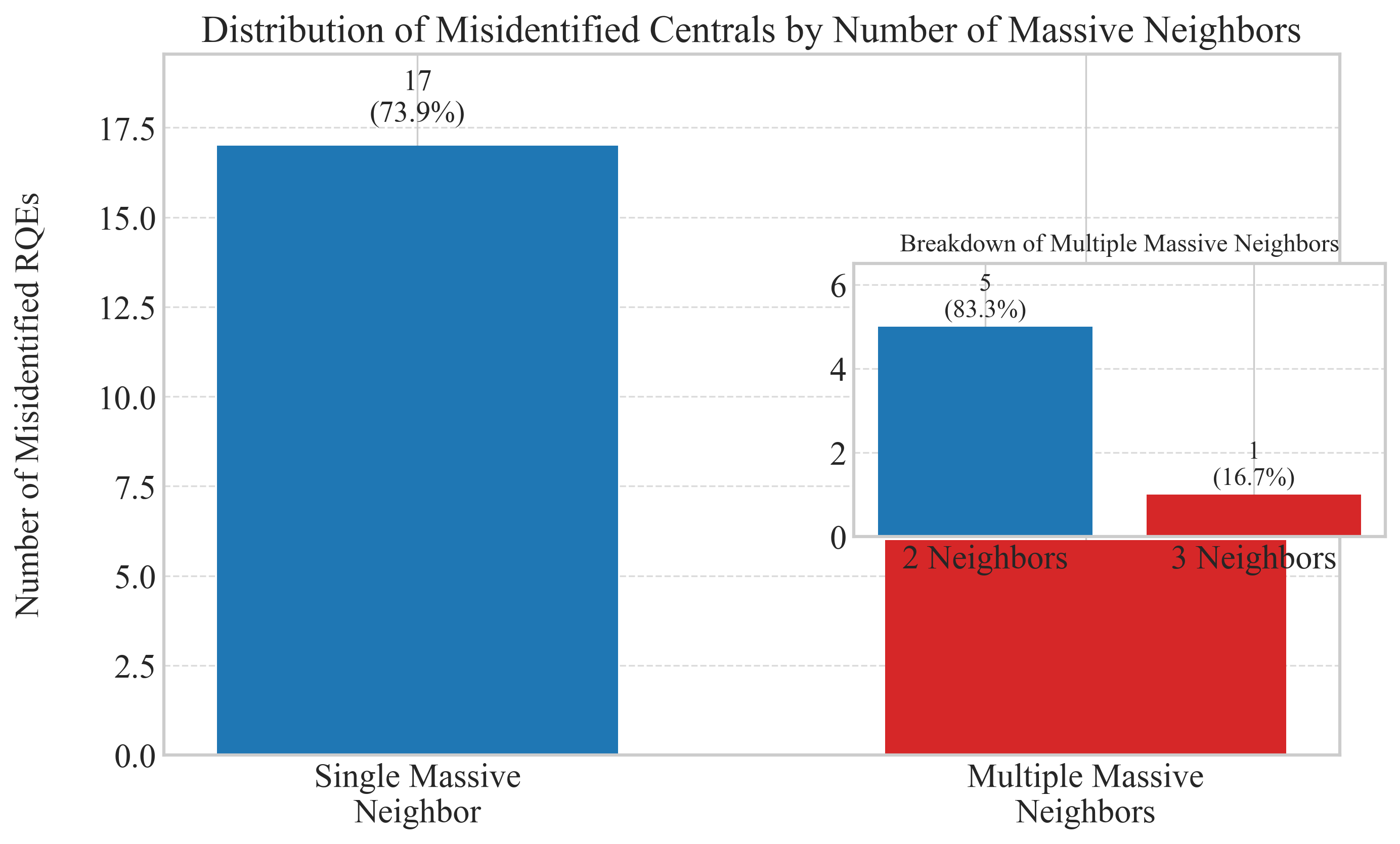}\\
        \vspace{2pt}
        {\small\textbf{(b)} Number of massive neighbors among misidentified centrals.}
    \end{minipage}

    % Second row: One panel spanning full width
    \vspace{0.4cm}
    \begin{minipage}[t]{\textwidth}
        \centering
        \includegraphics[width=0.75\textwidth]{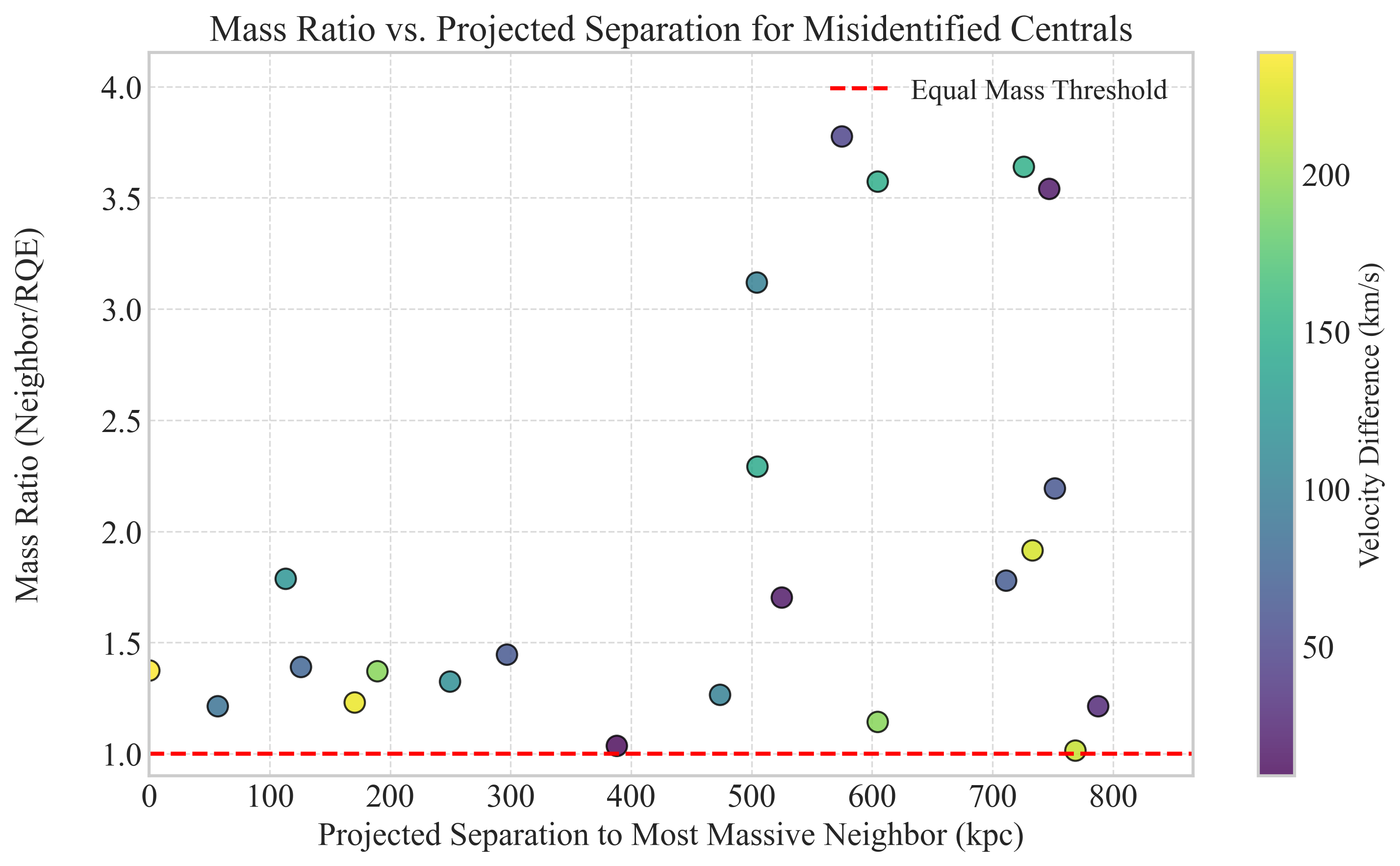}\\
        \vspace{2pt}
        {\small\textbf{(c)} Stellar mass ratio vs. projected separation for the most massive neighbor of each misidentified central. Color bar indicates velocity separation.}
    \end{minipage}

    \caption{Centrality reassessment for RQEs. Panel (a) shows how many RQEs remain true centrals or are reclassified as misidentified centrals under our criteria. Panel (b) breaks down the number of massive neighbors among the misidentified centrals. Panel (c) shows how close and how much more massive these neighbors are, with velocity separation color-coded.}
    \label{fig:centrality_reassessment}
\end{figure*}

Using this method, we find that \textbf{132 RQEs (85.2\%)} qualify as true centrals, while the remaining \textbf{23 RQEs (14.8\%)} have more massive neighboring galaxies within their local group environment and are thus reclassified as misidentified centrals (Figure~\ref{fig:centrality_reassessment}a). 
Among the misidentified centrals, \textbf{17 RQEs (73.9\%)} have a single more massive neighbor, while \textbf{6 RQEs (26.1\%)} have multiple massive neighbors (Figure~\ref{fig:centrality_reassessment}b). In the multiple-neighbor subset, \textbf{5 systems} have two massive neighbors, and \textbf{1 system} has three. These systems represent complex environments with ambiguous central-satellite dynamics, where multiple galaxies of comparable mass are located within close proximity, complicating the group hierarchy.

To better understand the nature of these reclassifications, we examine how massive and how close these more massive neighbors are relative to the RQEs. Figure~\ref{fig:centrality_reassessment}c shows the stellar mass ratio (neighbor/RQE) versus the projected physical separation for the most massive neighbor of each misidentified central RQE. The color bar encodes the line-of-sight velocity difference between the RQE and its more massive companion.

We find that in \textbf{30\%} of the misidentified centrals, the most massive neighbor is at least \textbf{twice as massive} (i.e., mass ratio $>2$), and interestingly, all such neighbors are located beyond \textbf{500 kpc} from the RQE but remain within the virial radius typically associated with galaxy groups. 

Together, these findings indicate that while the majority of RQEs are indeed the dominant galaxies in their halos, a significant minority are potentially satellites misclassified as centrals. This has important implications for interpreting their evolutionary histories and inferring the role of internal versus external quenching processes. 

\subsubsection{Isolation Status of True Central RQEs}

Having reclassified the centrality status of RQEs in the previous section, we now assess the degree of isolation among the 132 RQEs identified as \textbf{true centrals}. This analysis is crucial for disentangling the roles of internal versus external processes in galaxy quenching. Truly isolated central galaxies$-$those not subject to significant gravitational influence from nearby companions$-$are less likely to experience quenching via environmental mechanisms such as tidal stripping or gas removal, and more likely to be affected by internal processes alone.

We apply the unified isolation criterion defined in Section~\ref{sec:isolation_method}, which classifies a central RQE as \textbf{isolated} if it has no neighboring galaxies with stellar mass $M_{\text{neighbor}} \geq 0.1 \times M_{\text{RQE}}$ within a projected distance $R_{\text{proj}} \leq 0.5$ Mpc and line-of-sight velocity difference $\Delta V \leq 200$ km s$^{-1}$. Neighbors satisfying these criteria are deemed \textit{influential}, due to their potential to induce structural or dynamical perturbations, minor mergers, or gas exchange.

\begin{figure*}[ht!]
    \centering
    \includegraphics[width=0.75\textwidth]{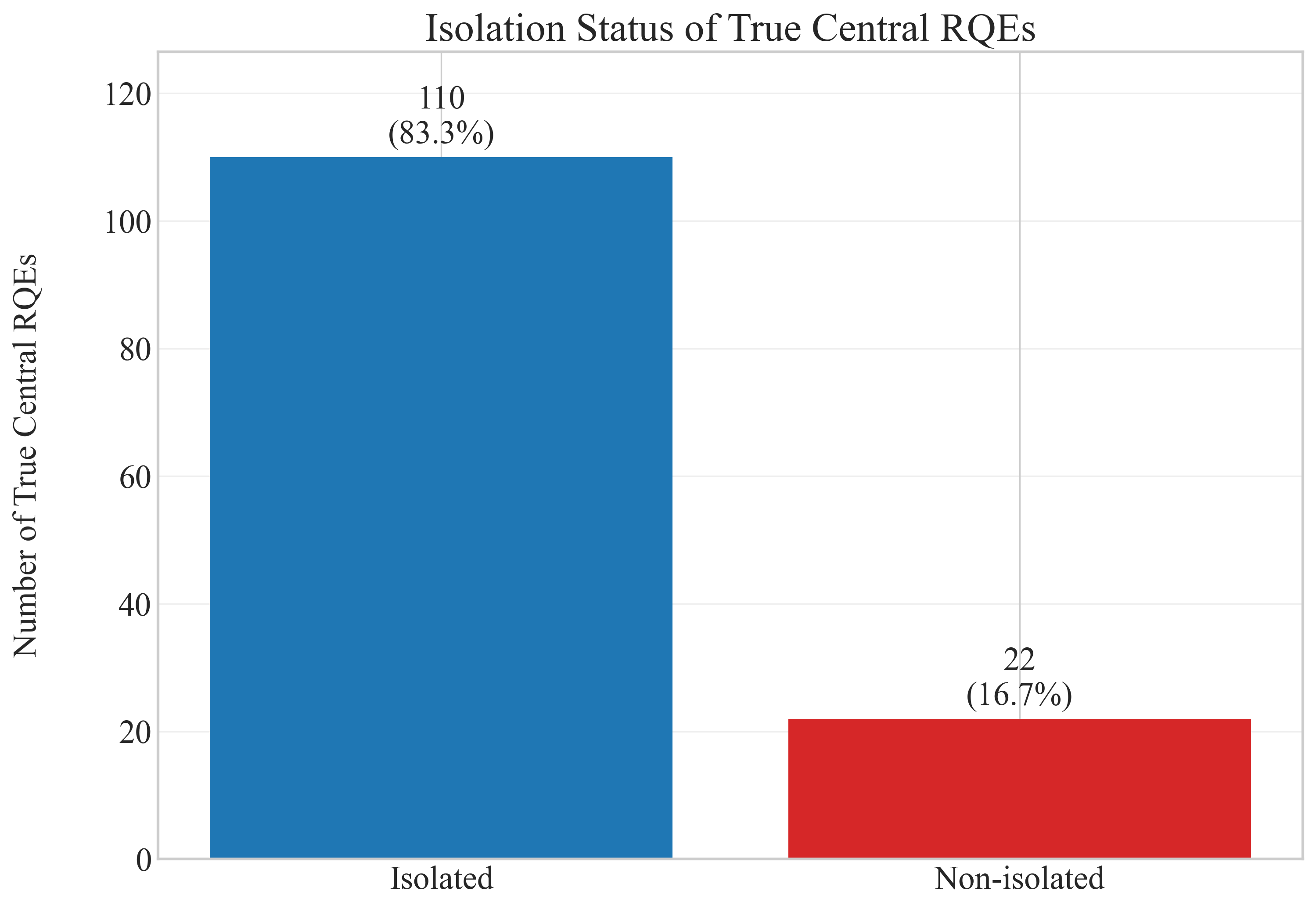}\\
    \vspace{2pt}
    {\small\textbf{(a)} Number of true central RQEs classified as isolated or non-isolated using our physically motivated isolation criterion.}

    \vspace{0.5cm}
    
    \includegraphics[width=0.75\textwidth]{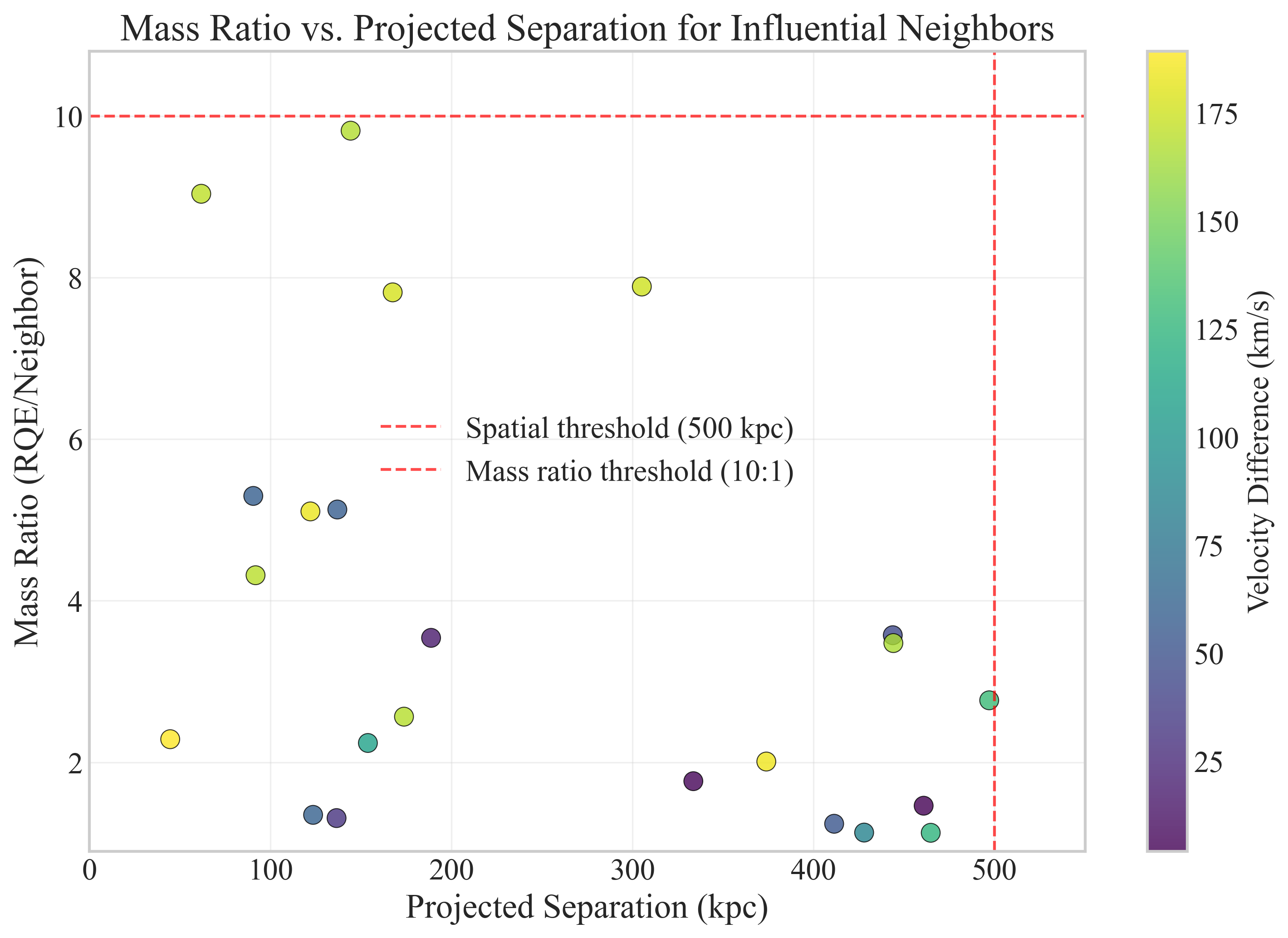}\\
    \vspace{2pt}
    {\small\textbf{(b)} Distribution of influential neighbors in terms of projected separation and mass ratio. Color bar indicates line-of-sight velocity separation.}
    
    \caption{Isolation properties of true central RQEs based on our revised definition. Panel (a) compares isolated and non-isolated populations, and Panel (b) characterizes the gravitational influence of nearby companions.}
    \label{fig:isolation_diagnostics}
\end{figure*}

Applying this definition, we find that \textbf{110 out of 132 true centrals (83.3\%)} are classified as isolated, while \textbf{22 RQEs (16.7\%)} are \textbf{non-isolated}, as they host at least one influential neighbor (Figure~\ref{fig:isolation_diagnostics}a). Among these non-isolated systems, the majority (\textbf{21 out of 22}) have only one influential neighbor, while one RQE has two.

To further investigate the characteristics of the local environments around non-isolated RQEs, we examine the properties of their influential neighbors. Figure~\ref{fig:isolation_diagnostics}b shows the projected separation ($R_{\text{proj}}$) versus stellar mass ratio (RQE/neighbor) for each influential neighbor. The color bar encodes the velocity offset $\Delta V$ between the RQE and its neighbor. Most influential neighbors are found either within 200 kpc or in the range of 300--500 kpc, with the majority having stellar mass ratios between 1:1 and 4:1. This suggests that even though these companions are not more massive than the RQEs, their relative proximity and mass can still render them significant in shaping RQE evolution through flybys or minor interactions.

When comparing our isolation classifications to those from the Y07 group catalog, we find notable differences. Among the 132 true central RQEs, Y07 classified 99 (75.0\%) as isolated. Under our physically motivated criteria, this proportion increased to 110 RQEs (83.3\%), indicating a net increase in isolated systems. However, the individual reclassifications reveal a more complex pattern: only 9 RQEs (9.1\% of Y07's isolated sample) were reclassified as non-isolated, while 20 RQEs (60.6\% of Y07's non-isolated sample) were reclassified as isolated. This pattern demonstrates that Y07's group-finding algorithm was generally more conservative in assigning isolation status, particularly for RQEs in sparse environments that our criteria recognize as truly isolated.

\subsubsection{Identification of Pseudo-Central RQEs}

Beyond establishing the isolation status of true central RQEs, we now investigate their potential association with massive cluster-scale halos. This analysis addresses whether some RQEs, despite qualifying as central galaxies within their immediate group environment, may be influenced by or infalling into larger structures$-$a scenario that could significantly impact their quenching mechanisms.

We define \textbf{pseudo-central RQEs} as systems that meet our criteria for true centrals (Section~\ref{method2}) but are located within the sphere of influence of a nearby massive cluster. Specifically, an RQE is classified as a pseudo-central if there exists a cluster-sized halo ($M_{\text{halo}} \geq 10^{14} \, M_{\odot}$) within $R_{\text{proj}} \leq 2$ Mpc and $\Delta V \leq 2000$ km s$^{-1}$.

\begin{figure*}[ht!]
    \centering
    \includegraphics[width=\textwidth]{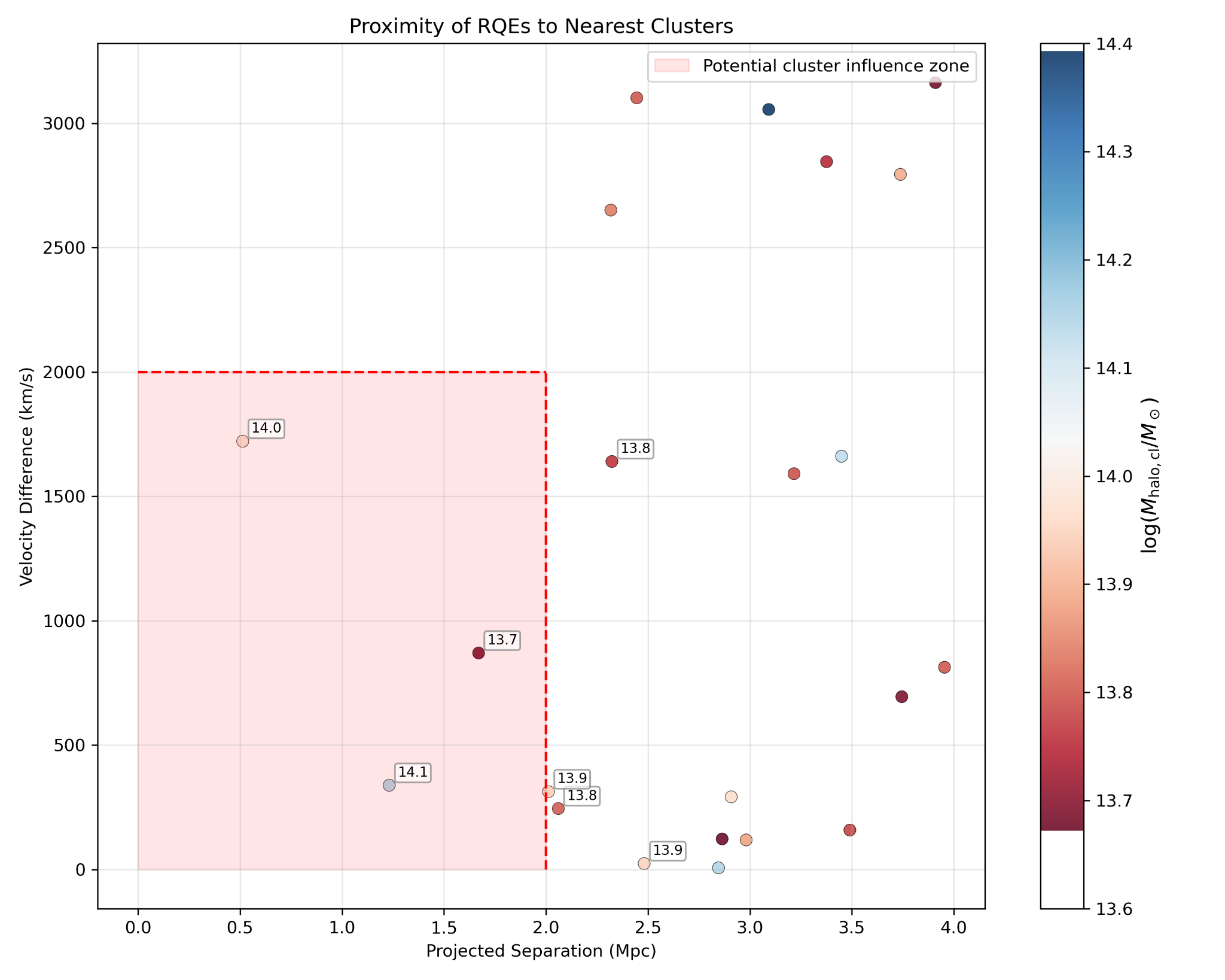}
    \caption{Proximity of RQEs to their nearest galaxy clusters in phase space. The shaded region represents our pseudo-central criteria ($R_{\text{proj}} \leq 2$ Mpc and $\Delta V \leq 2000$ km s$^{-1}$). Points are color-coded by the logarithmic halo mass of the cluster. Text labels indicate $\log(M_{\text{halo}}/M_{\odot})$ values for clusters within or near the influence zone. Only one true central RQE qualifies as a pseudo-central by our strict criteria, with a massive cluster ($\log(M_{\text{halo}}/M_{\odot}) = 14.1$) within the influence zone.}
    \label{fig:pseudo_central}
\end{figure*}

Figure~\ref{fig:pseudo_central} presents the distribution of RQEs relative to their nearest clusters, based on a crossmatch with the \citet{2024ApJS..272...39W} cluster catalog. Each point represents the nearest cluster to an RQE within a broader search window of $R_{\text{proj}} \leq 4$ Mpc and $\Delta V \leq 4000$ km s$^{-1}$. The shaded region delineates our pseudo-central criteria, and point colors indicate the logarithmic halo mass of the cluster. In total, only 22 RQEs were found to have at least one cluster within this broader search radius, suggesting that the majority of RQEs in our sample inhabit relatively low-density environments.

Our cross-matching with the \citet{2024ApJS..272...39W} cluster catalog reveals that only a single true central RQE qualifies as a pseudo-central by our strict definition. This RQE has a nearby cluster with $\log(M_{\text{halo}}/M_{\odot}) = 14.1$ at a projected separation of approximately 1.2 Mpc and a velocity difference of about 340 km/s, placing it well within our influence zone. Interestingly, we identified two additional RQEs with massive clusters ($\log(M_{\text{halo}}/M_{\odot}) \approx 14.0$) just beyond our 2 Mpc threshold but with relatively low velocity separations ($<$ 500 km/s), suggesting that these systems might also experience some degree of cluster influence.

When examining the distribution more broadly, we note that the remaining RQE-cluster pairs in our influence zone involve either less massive clusters ($\log(M_{\text{halo}}/M_{\odot}) < 13.8$) or RQEs previously classified as misidentified centrals in our analysis. This remarkably low incidence of pseudo-centrals (approximately 0.8\% of our true central sample) suggests that most true RQE centrals has evolved largely independent of direct cluster-scale environmental influences and are not currently undergoing strong cluster-driven preprocessing or infall, and instead reside predominantly in field or group-scale environments.

\subsubsection{Quantifying the Local Environment of RQEs}

With the centrality, isolation, and pseudo-central statuses of RQEs now established, we conclude our results section by quantifying their local environments using two physically motivated density metrics: the \textit{projected local number density} ($\sigma_n$) and the \textit{surface stellar mass density} ($\Sigma_*$). These complementary measures help characterize both the richness and gravitational potential of an RQE's surrounding environment and allow us to distinguish between environments where external quenching mechanisms may be relevant versus those likely dominated by internal processes.

Following the methodology outlined in Section~\ref{sec:envmetrics}, we compute $\sigma_n$ as the number of galaxies within a projected aperture of 0.8 Mpc and a line-of-sight velocity window of $\Delta V \leq 250$ km s$^{-1}$, divided by the area of the aperture. The surface stellar mass density $\Sigma_*$ is calculated by summing the stellar masses of all galaxies (including the RQE itself) within the same spatial and velocity limits and dividing by the projected area. If a given RQE has no neighbors within this volume, it contributes to the minimum possible density values: $\sigma_n \approx 0.50$ Mpc$^{-2}$ and $\Sigma_*$ approximately equal to the stellar mass of the RQE itself divided by the aperture area.

\begin{figure*}[ht!]
    \centering
    \includegraphics[width=\textwidth]{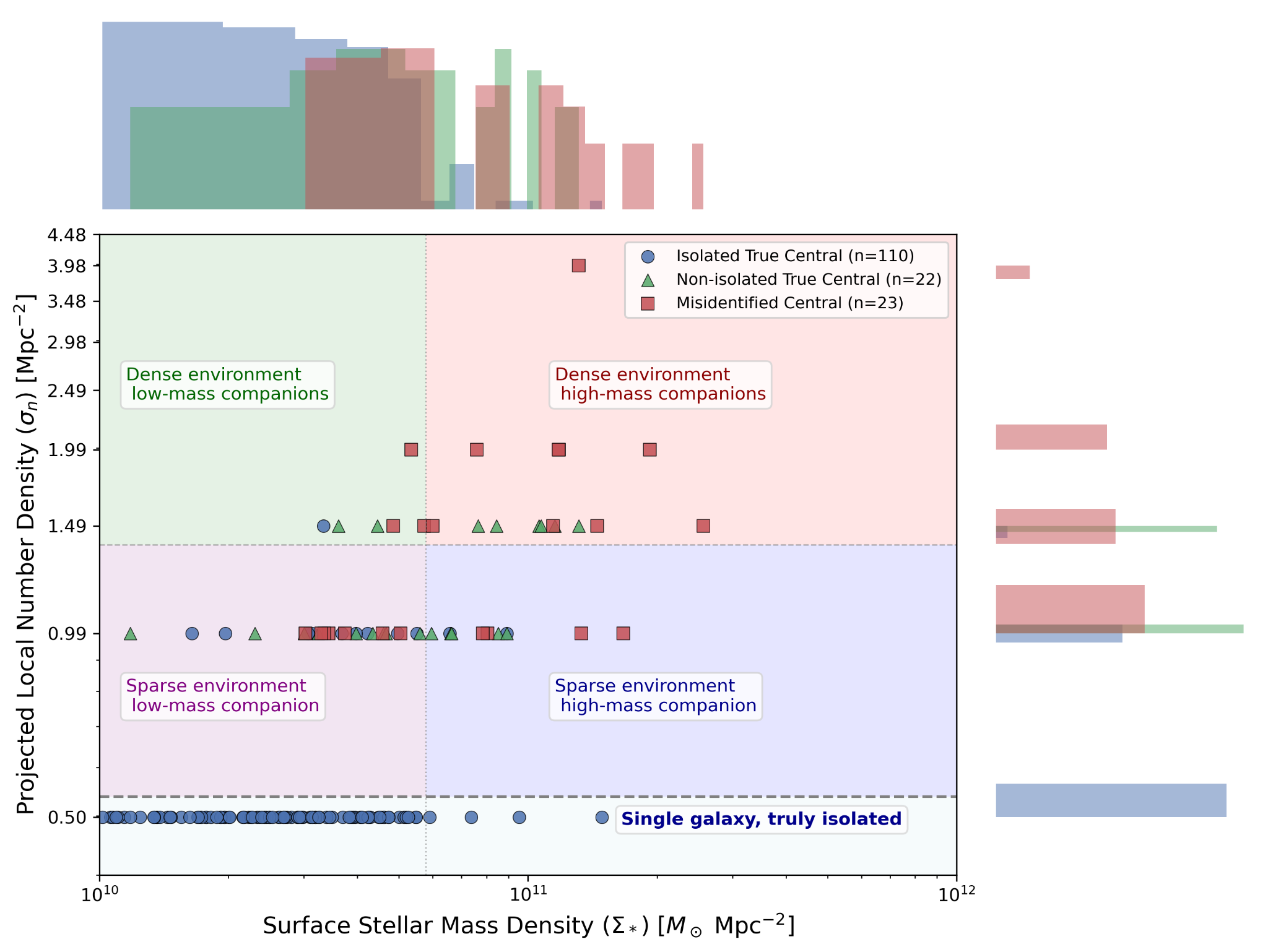}
    \caption{Environmental metrics for different RQE categories. The main plot shows the relationship between projected local number density ($\sigma_n$) and surface stellar mass density ($\Sigma_*$) for all 155 RQEs in our sample, with points color-coded by category. The horizontal dashed lines mark the densities for a single galaxy (0.50 galaxies Mpc$^{-2}$) and the boundary between environments with single versus multiple neighbors (1.39 galaxies Mpc$^{-2}$). The vertical dotted line indicates the median mass density of non-isolated true central RQEs. These boundaries define four distinct environmental regimes, as labeled. Side panels show the distributions of each metric for the three RQE categories.}
    \label{fig:env_metrics}
\end{figure*}

Figure~\ref{fig:env_metrics} shows the distribution of 155 RQEs in the $\Sigma_* - \sigma_n$ plane, separated by central classification. Most \textbf{isolated true centrals} (shown as blue circles) cluster along the single-galaxy line ($\sigma_n \approx 0.50$ Mpc$^{-2}$), indicating the absence of any significant companions within their immediate environments. By contrast, both \textbf{non-isolated true centrals} (green triangles) and \textbf{misidentified centrals} (orange squares) exhibit systematically higher values of both number and stellar mass density, consistent with denser and more complex environments.

To better interpret environmental regimes, we divide the density space into quadrants based on two thresholds:
\begin{itemize}
    \item A vertical line at the \textbf{median $\Sigma_*$ value of non-isolated true centrals}, separating systems dominated by massive neighbors from those surrounded by lower-mass galaxies.
    \item A horizontal line at $\sigma_n \approx 1.39$ Mpc$^{-2}$, which corresponds to the transition between environments with a single neighbor versus multiple neighbors. This boundary distinguishes between sparse environments (at most one companion) and denser environments with multiple potential interactions.
\end{itemize}

These boundaries define four quadrants:
\begin{enumerate}
    \item \textbf{Dense environment, high mass companions} (high $\sigma_n$, high $\Sigma_{\ast}$): Characterized by environments containing multiple, massive neighbors. RQEs in this regime are likely subject to strong gravitational interactions. This quadrant predominantly hosts misidentified centrals and non-isolated true centrals.
    
    \item \textbf{Dense environment, low-mass companions} (high $\sigma_n$, low $\Sigma_{\ast}$): Comprising systems surrounded primarily by numerous but less massive neighbors; could promote weak galaxy harassment or interactions.
    
    \item \textbf{Sparse environment, high-mass companion} (low $\sigma_n$, high $\Sigma_{\ast}$): 
    Defined by relatively isolated environments that nevertheless feature at most one high mass galaxy that contribute significantly to the local mass density. Despite overall sparsity, gravitational interactions from the massive neighbor can be significant affecting the star formation activity.
       
    \item \textbf{Sparse environment, low-mass companion} (low $\sigma_n$, low $\Sigma_{\ast}$): This regime represents sparse environments with minimal external gravitational influence. These systems have at most one neighbor, which contributes relatively little to the total environmental mass.
\end{enumerate}

The distribution of RQE subpopulations across these quadrants reveals notable trends. As expected, misidentified centrals tend to populate the denser, high-mass regions. However, a small subset (4 galaxies) appears in the sparse, low-mass companion quadrant. These special cases likely represent borderline systems where the RQE has a single neighbor that is more massive than itself (thus triggering the ``misidentified central" classification) but not massive enough to push the environmental surface mass density above our threshold. 

Notably, all 72 RQEs lying on the single-galaxy density line are classified as isolated true centrals, reinforcing the notion that a significant portion of RQEs are found in environments with negligible external influence. For these systems, any cessation of star formation is likely attributed to internal quenching mechanisms.

The median density values for each subpopulation further support these interpretations. Isolated true centrals have typical number and mass densities of 0.50 galaxies Mpc$^{-2}$ and $2.85 \times 10^{10}$ M$_{\odot}$ Mpc$^{-2}$, respectively. Non-isolated true centrals show elevated values of 0.99 galaxies Mpc$^{-2}$ and $5.77 \times 10^{10}$ M$_{\odot}$ Mpc$^{-2}$, while misidentified centrals occupy the highest density regime with 1.49 galaxies Mpc$^{-2}$ and $7.60 \times 10^{10}$ M$_{\odot}$ Mpc$^{-2}$.

These results reinforce that while the majority of RQEs occupy genuinely isolated, low-density environments, a diverse range of environmental conditions exists. This environmental heterogeneity points to multiple evolutionary pathways for RQEs, with internal processes dominating in truly isolated systems and external environmental mechanisms playing a larger role in richer environments.

\section{Discussion}\label{disc2}

Our environmental reassessment of Recently Quenched Ellipticals (RQEs) reveals important insights into the physical conditions that govern quenching in these systems. By systematically reevaluating their group membership, centrality status, and local environmental context, we uncover a more nuanced understanding of their evolutionary pathways than previously available.

A key result of this analysis is that approximately 29\% (45 out of 155) of RQEs originally identified as central galaxies in the Yang et al. (2007) group catalog are more likely to be environmentally influenced systems. Of these, 23 are now classified as misidentified centrals$-$likely massive satellites residing near more massive neighbors$-$while the remaining 22 are true centrals that reside in non-isolated environments. This reassignment is based on the presence of at least one more massive galaxy within a projected radius of $R_{\text{proj}} \leq 0.8$~Mpc and a velocity separation of $\Delta V \leq 250$~km~s$^{-1}$, criteria that suggest these galaxies are not dynamically dominant within their local potential wells.

This misidentification rate aligns with previous studies demonstrating that central galaxy identification in group catalogs can be incorrect in 25\% of low-mass halos and up to 40\% in massive halos ($>5\times 10^{13} M_{\odot}$) \cite{2011MNRAS.410..417S}.
Such misclassifications can introduce significant biases in quenching studies, particularly when satellite-specific processes (e.g., ram-pressure stripping, strangulation) are incorrectly attributed to quenching in central galaxies.
Our improved centrality reassessment thus provides a more robust framework to disentangle the competing roles of internal versus environmental quenching processes.

\subsection{Disentangling Intrinsic and Environmental Quenching Channels}

Building on our revised classification framework, we divide the RQE population into two broad evolutionary regimes:

\begin{itemize}
    \item \textbf{Intrinsic regime} $-$ 110 RQEs (71\%) identified as \textit{isolated true centrals}, with no evidence of nearby massive companions.
    \item \textbf{Environmental regime} $-$ 45 RQEs (29\%) classified as either \textit{non-isolated true centrals} or \textit{misidentified centrals}, likely affected by group-scale interactions.
\end{itemize}

RQEs in the intrinsic regime are expected to have quenched solely via internal processes, whereas those in the environmental regime may have their star formation activity influenced through interactions with their surroundings.

\begin{figure*}
    \centering
    \includegraphics[width=\textwidth]{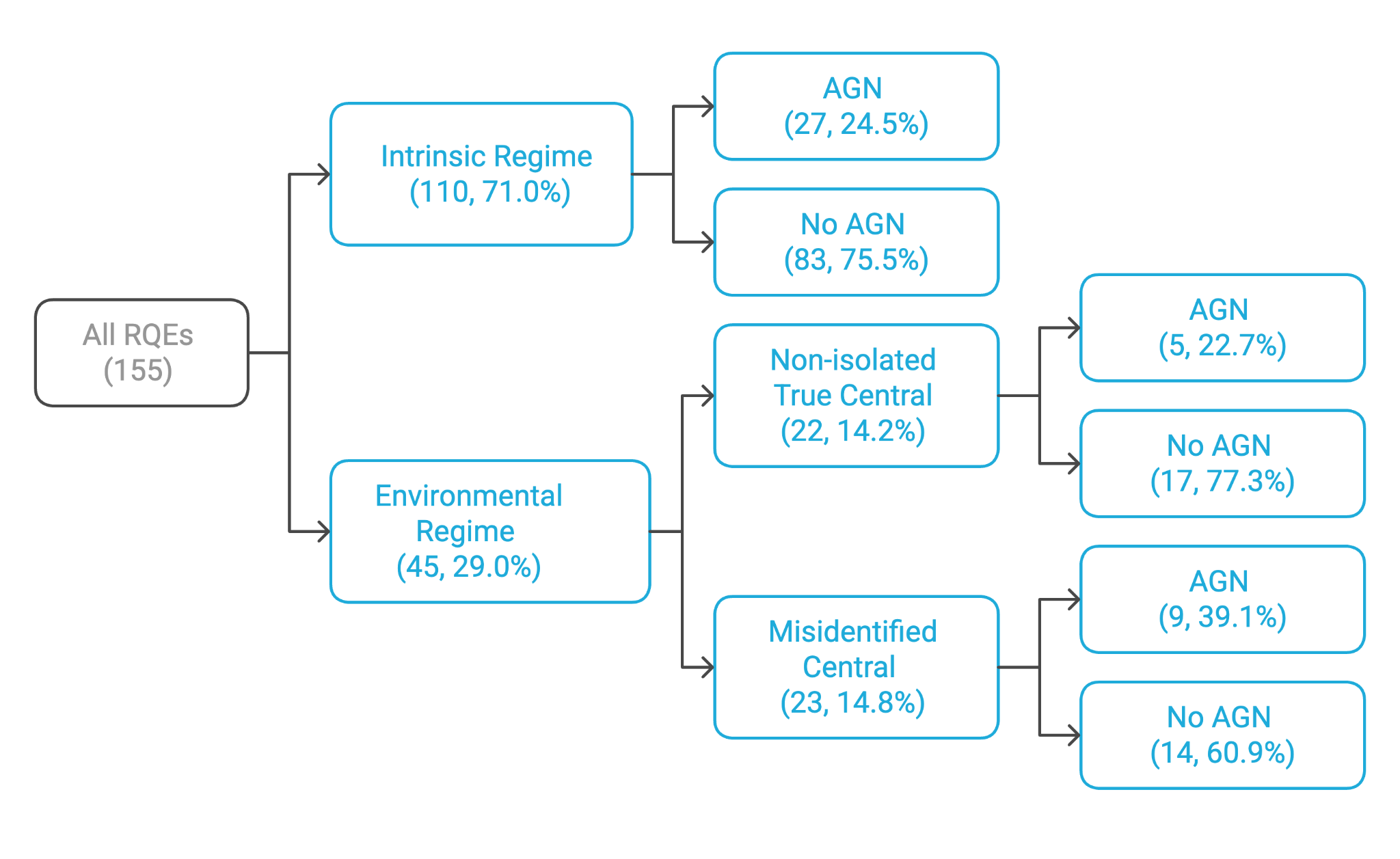}
    \caption{Hierarchical breakdown of RQEs by quenching regime and AGN presence.}
    \label{flowchart_ch2}
\end{figure*}

To explore this further, we subdivide each regime by emission-line classification. RQEs with Seyfert- or LINER-like line ratios are grouped into an \textbf{AGN-like} category, while those without such signatures are labeled \textbf{non-AGN}. We note that LINER-like line ratios do not uniquely imply black-hole accretion and can also arise from other ionization sources; we therefore treat this classification as an empirical spectroscopic category rather than a definitive AGN identification. This distinction serves as a proxy for the presence or absence of internal energetic feedback that might contribute to quenching. Table~\ref{tab:rqe_quenching_summary} and Figure~\ref{flowchart_ch2} summarize the breakdown of our sample by quenching regime and emission-line category.

\begin{deluxetable*}{lccc}
\tablecaption{Classification of RQEs by Quenching Regime and AGN Presence \label{tab:rqe_quenching_summary}}
\tablewidth{0pt}
\tablehead{
\colhead{Category} & \colhead{Count} & \colhead{Percentage (\%)} & \colhead{AGN (Seyfert/LINER)}
}
\startdata
Total RQEs & 155 & 100 & \nodata \\
\hline
\textbf{Intrinsic Regime} (Isolated True Centrals) & 110 & 71.0 & 27 (24.5\%) \\
\textbf{Environmental Regime} & 45 & 29.0 & 14 (31.1\%) \\
\hspace{1em}Non-isolated True Centrals & 22 & 14.2 & 5 \\
\hspace{1em}Misidentified Centrals & 23 & 14.8 & 9 \\
\enddata
\end{deluxetable*}

Among the 41 AGN-hosting RQEs, 27 (65.8\%) reside in the intrinsic regime. This supports the idea that AGN feedback is a viable mechanism for sustaining quenching in these systems. These galaxies typically retain moderate HI reservoirs, consistent with scenarios in which gas is stabilized or heated---rather than expelled---thus preventing further star formation.

The environmental regime displays a more diverse set of quenching pathways. Of the 45 environmentally influenced RQEs, 14 (31.1\%) are AGN hosts, while the remainder are optically quiescent. Within this group, \textbf{misidentified centrals}$-$defined as RQEs with at least one more massive neighbor within a projected distance of 0.8 Mpc and a velocity offset of $\leq$ 250 km s$^{-1}$$-$are likely to be massive satellites. These galaxies may be experiencing quenching via satellite-specific environmental processes such as strangulation, ram-pressure stripping, or tidal interactions \cite{2012MNRAS.424..232W, 2015Natur.521..192P}.

\begin{figure*}
    \centering
    \includegraphics[width=\textwidth]{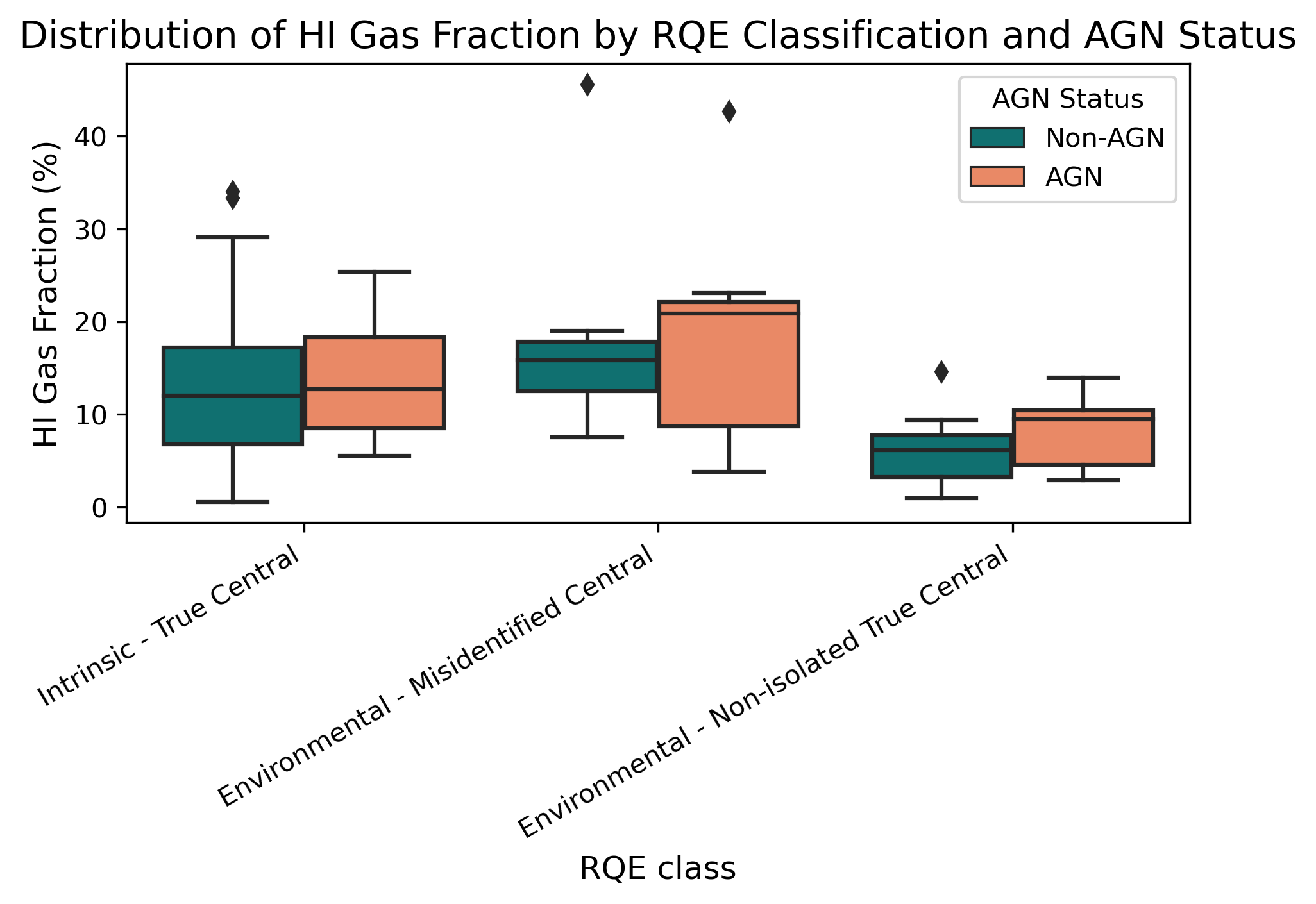}
    \caption{Distribution of HI gas fractions (in percent) across RQEs classified by quenching regime and AGN presence. The plot highlights systematic differences between intrinsic and environmental RQEs, with non-isolated true centrals showing the lowest median gas fractions irrespective of AGN presence.}
    \label{HIgas_quench}
\end{figure*}

Interestingly, misidentified centrals exhibit the highest median inferred HI gas fraction (18.1\%) among all RQE subtypes, rising to 22\% for those with AGN emission. This is notably higher than the median values observed for non-isolated true centrals (8.5\%) and isolated true centrals (12.0\%). These differences are illustrated in Figure~\ref{HIgas_quench}, which compares HI gas fractions across quenching regimes and AGN classifications. The finding that massive satellites show evidence for non-negligible HI while remaining quenched suggests one of two possibilities: either these galaxies are in an early phase of environmental quenching$-$where star formation has recently halted but gas removal is still ongoing$-$or they represent a distinct evolutionary pathway in which the gas remains abundant but is inefficient at forming stars, possibly due to stabilization by internal or environmental processes.

The latter scenario is particularly compelling for AGN-hosting misidentified centrals, where nuclear feedback may work synergistically with environmental effects to maintain quiescence without complete gas removal. This combination of mechanisms $-$ nuclear heating and environmental pressure$-$could explain why these systems exhibit both quenched star formation and elevated gas fractions. The environmental influence likely contributes to triggering or sustaining AGN activity through gas displacement toward the central regions, while the AGN in turn helps maintain the quenched state by preventing the gas from cooling efficiently enough to form stars. This interplay between environmental and internal processes creates a stable configuration where substantial HI reserves can coexist with a quiescent stellar population, representing a unique transitional phase in satellite galaxy evolution.

A more puzzling subset of RQEs in the environmental regime is the population of \textbf{non-isolated true centrals}---central galaxies with nearby influential neighbors. Despite their group-centric positions and possible exposure to gas-rich interactions via minor mergers, they exhibit the \textit{lowest} median HI gas fractions (8.5\%) among all RQE classes, and the \textit{oldest} median light-weighted stellar ages ($\sim$2.21 Gyr). 
These properties suggest that these systems are not in a rejuvenating phase. While proximity to massive companions might, in principle, allow for minor gas accretion, the observed lack of substantial HI reservoirs implies that such accretion has either not occurred or is not efficient enough to overcome environmental suppression.
Furthermore, any residual HI is likely too diffuse or too warm to meet the critical thresholds for star formation, such as the required column densities ($N_{\text{HI}} \gtrsim 10^{20.5}~\text{cm}^{-2}$) and temperatures below $\sim 20$K. These systems are thus more consistent with a scenario of slow, environmentally driven quenching---such as strangulation---where cold gas reservoirs are gradually exhausted or stabilized, leaving behind galaxies that are quenched but retain low-level HI content insufficient to trigger renewed star formation.

Among the isolated true centrals in the intrinsic regime, 24.5\% exhibit AGN-like emission, reinforcing the role of internal feedback in maintaining quiescence. The coexistence of HI gas and AGN activity in these galaxies suggests a feedback loop in which the gas is dynamically stabilized or thermally suppressed, thereby inhibiting further star formation without fully depleting the cold gas reservoir.

The remaining 83 isolated true centrals without AGN emission represent the most clearly intrinsically quenched RQEs. In the absence of both external environmental effects and AGN-like emission, their sustained quiescence likely results from internal processes. These may include morphological quenching, wherein stellar bulges stabilize the gas disk against fragmentation, or virial shock heating in halos near $10^{12}~M_\odot$, which can suppress the cooling and accretion of new gas. While these galaxies do retain modest HI reservoirs, this alone is not sufficient to trigger star formation. The inefficiency may reflect gas that is either too diffuse or dynamically stabilized. Follow-up investigations---such as measuring HI surface densities or tracing molecular gas content---are needed to determine whether these galaxies are unable to form stars due to physical stability thresholds (e.g., elevated Toomre $Q$), or whether they are simply awaiting future gas accretion.

Finally, we find that pseudo-centrals---true centrals located within the influence region of a nearby massive cluster---are extremely rare in our sample (only one system by our adopted thresholds). This indicates that, for RQE centrals in this sample, direct cluster-scale preprocessing is unlikely to be a dominant pathway, and that the relevant environmental effects (when present) are more plausibly group-scale or companion-driven.

Taken together, these findings suggest that \textbf{intrinsic quenching mechanisms dominate the RQE population}, particularly among isolated true centrals. However, external factors---especially in the form of satellite-like conditions or dense local environments---also play a significant role in shaping the fate of a substantial minority. Furthermore, the presence of AGN-like emission in both regimes underscores that internal and environmental processes are not mutually exclusive but may instead act in tandem to suppress or regulate star formation.

These results support a dual-pathway model for RQE evolution: roughly 70\% appear to be quenched through internal processes, often aided by nuclear feedback, while the remaining 30\% show signatures of environmental influence, including group-scale interactions and satellite-like quenching. The coexistence of HI reservoirs and quiescent stellar populations in many of these systems highlights the complex interplay between gas content, feedback, and environment in the transition from star-forming to passive galaxies.

Recently quenched ellipticals, therefore, are not a monolithic class but rather a diverse population occupying multiple evolutionary tracks---some shaped by internal feedback, others by external environmental processes, and many by a combination of both.

% \subsection{Caveats and Future Work}

% While our results are robust, several limitations remain. Our analysis is limited to the SDSS footprint and redshift depth, which may bias against detecting very low-mass companions or extended cluster infall regions. Spectroscopic incompleteness and fiber collisions in SDSS also reduce completeness in dense regions.

% Future work could incorporate deeper photometric data, higher-resolution HI mapping, or IFU surveys such as MaNGA to further dissect the role of internal versus external quenching. Moreover, studying the gas content and kinematics of isolated versus non-isolated RQEs could reveal whether minor interactions are sufficient to fuel or suppress star formation in these transitional systems.

\section{Conclusion}\label{conc2}

In this study, we revisited the environmental context of Recently Quenched Ellipticals (RQEs) to assess the relative contributions of internal and external mechanisms in driving their quenching. Using revised classifications based on physically motivated criteria for centrality and isolation, we found that the RQE population exhibits considerable diversity in environmental properties, gas content, and feedback signatures.

A primary finding of this investigation is that the majority of RQEs (71\%) exist as isolated true central galaxies that have likely quenched through predominantly internal mechanisms. The scarcity of pseudo-centrals (only one RQE) further emphasizes that large-scale cluster environments play a minimal role in RQE formation. This challenges traditional paradigms that emphasize environmental processes as the primary drivers of morphological transformation and quenching in early-type galaxies.

Nevertheless, a significant minority (29\%) of RQEs show clear evidence of environmental influence, with 14.8\% reclassified as misidentified centrals (likely massive satellites) and 14.2\% identified as non-isolated true centrals. This environmental diversity, coupled with varying levels of AGN activity across different RQE subpopulations, points to multiple evolutionary pathways rather than a single universal quenching mechanism.

The surprising variations in HI gas content among RQE subgroups provide critical clues about their evolutionary states. Misidentified centrals appear to have high inferred HI gas fractions (18.1\%, rising to 22\% for those with AGN) on average than other subgroups, suggesting they are in early stages of environmental processing where star formation has recently halted but gas removal remains incomplete. Conversely, non-isolated true centrals exhibit the lowest inferred HI gas fractions (8.5\%) despite their central status, indicating prolonged environmental interactions that have gradually depleted their gas reservoirs without triggering significant rejuvenation.
Perhaps most intriguing are the 83 non-AGN isolated true centrals, which remain quiescent despite substantial gas reserves and minimal external influence. Their continued quiescence likely stems from internal structural properties$-$such as bulge-stabilized gas disks or virial shock heating$-$that render their gas inefficient at forming stars despite its abundance.

These findings support a dual-pathway framework for RQE evolution:

\begin{itemize}
    \item \textbf{Internal quenching pathway}: Dominated by isolated true centrals (71\%), where mergers trigger morphological transformation followed by quenching through AGN feedback (24.5\%) or morphological stabilization (75.5\%), maintaining quiescence while preserving substantial gas reservoirs.
    
    \item \textbf{Environmental influence pathway}: Comprising non-isolated true centrals and misidentified centrals (29\%), where external interactions contribute significantly to quenching, though with notably different outcomes for gas content and star formation potential.
\end{itemize}

The presence of AGN/LINER activity at similar rates across both pathways (24.5\% in isolated systems, 31.1\% in environmental systems) suggests that nuclear processes operate largely independently of environmental density while remaining important quenching mechanisms in both contexts.

Our results demonstrate that RQEs represent an important transitional stage in galaxy evolution where diverse physical processes$-$both internal and environmental$-$contribute to quenching while often allowing significant HI gas to remain. This challenges simplistic gas depletion models and highlights the complexity of pathways through which galaxies transition from star-forming to quiescent states.

Future investigations should prioritize spatially resolved observations of both atomic (HI) and molecular (H$_2$) gas in RQEs to constrain the spatial distribution, surface densities, and kinematics of the cold gas reservoirs. Such measurements would offer critical diagnostics for quenching mechanisms like morphological stabilization and AGN-driven heating. Additionally, larger samples with comprehensive multi-wavelength coverage would enable more robust statistical analysis across the environmental spectrum, clarifying how various quenching processes interact and evolve over time.

\section*{Acknowledgments}
% We thank the anonymous referee for their valuable comments and suggestions, which have helped improve this paper. 
% The authors would like to thank the anonymous referee for their comments. 
DKD thanks Emily Matthews, Ruta Kale, Elizabeth Stoddard, Kalicharan Rajak, Vimla Devi, Aarti, Rajesh, Shriyanvi, and Paul Rulis for their help and support in completion of this paper.
This work made use of data from the Sloan Digital Sky Survey (SDSS), including the New York University Value-Added Galaxy Catalog (NYU-VAGC), and published value-added catalogs and group catalogs constructed from SDSS data. 
The SDSS Archive was created and distributed with the financial support of the Alfred P. Sloan Foundation, the Participating Institutions, the National Science Foundation, the U.S. Department of Energy, the National Aeronautics and Space Administration, the Japanese Monbukagakusho, the Max Planck Society, and the Higher Education Funding Council for England.
We also made use of a published massive-cluster catalog based on imaging from the DESI Legacy Imaging Surveys and \textit{WISE}. 
This research has made use of several software tools and services such as NASA's Astrophysics Data System Bibliographic Services, TOPCAT (Tools for OPerations on Catalogues And Tables, \cite{2005ASPC..347...29T}), Astropy\footnote{http://www.astropy.org},
a community developed core Python package for Astronomy \citep{2013A&A...558A..33A, 2018AJ....156..123A}, Scipy \citep{scipy}, Pandas \citep{pandas}, and Matplotlib \citep{2007CSE.....9...90H, matplotlib}.

\section*{Data Availability}
All input catalogs used in this work are publicly available through the literature sources cited in the text. The value-added catalogs and derived measurements generated for this study, which support the figures and results presented here, are available from the author upon reasonable request.

\bibliography{references}{}

@article{0004-637X-707-1-250,
  author={Marie Martig and FrÃ©dÃ©ric Bournaud and Romain Teyssier and Avishai Dekel},
  title={Morphological Quenching of Star Formation: Making Early-Type Galaxies Red},
  journal={The Astrophysical Journal},
  volume={707},
  number={1},
  pages={250},
  url={http://stacks.iop.org/0004-637X/707/i=1/a=250},
  year={2009}
}

@misc{deo2026investigatingquenchingrecentlyquenched,
      title={Investigating quenching in Recently Quenched Elliptical galaxies with HI studies}, 
      author={Deepak K. Deo and Daniel H. McIntosh and Sravani Vaddi and Kameswara B. Mantha and Ruta Kale and Alfonso G. Franco and Paul Rulis},
      year={2026},
      eprint={2601.08027},
      archivePrefix={arXiv},
      primaryClass={astro-ph.GA},
      url={https://arxiv.org/abs/2601.08027}, 
}

@INPROCEEDINGS{2005ASPC..347...29T,
       author = {{Taylor}, M.~B.},
        title = "{TOPCAT \& STIL: Starlink Table/VOTable Processing Software}",
    booktitle = {Astronomical Data Analysis Software and Systems XIV},
         year = 2005,
       editor = {{Shopbell}, P. and {Britton}, M. and {Ebert}, R.},
       series = {Astronomical Society of the Pacific Conference Series},
       volume = {347},
        month = dec,
        pages = {29},
       adsurl = {https://ui.adsabs.harvard.edu/abs/2005ASPC..347...29T}
}

@ARTICLE{2018AJ....156..123A,
       author = {{Astropy Collaboration} and {Price-Whelan}, A.~M. and {Sip{\H{o}}cz}, B.~M. and {G{\"u}nther}, H.~M. and {Lim}, P.~L. and {Crawford}, S.~M. and {Conseil}, S. and {Shupe}, D.~L. and {Craig}, M.~W. and {Dencheva}, N. and {Ginsburg}, A. and {VanderPlas}, J.~T. and {Bradley}, L.~D. and {P{\'e}rez-Su{\'a}rez}, D. and {de Val-Borro}, M. and {Aldcroft}, T.~L. and {Cruz}, K.~L. and {Robitaille}, T.~P. and {Tollerud}, E.~J. and {Ardelean}, C. and {Babej}, T. and {Bach}, Y.~P. and {Bachetti}, M. and {Bakanov}, A.~V. and {Bamford}, S.~P. and {Barentsen}, G. and {Barmby}, P. and {Baumbach}, A. and {Berry}, K.~L. and {Biscani}, F. and {Boquien}, M. and {Bostroem}, K.~A. and {Bouma}, L.~G. and {Brammer}, G.~B. and {Bray}, E.~M. and {Breytenbach}, H. and {Buddelmeijer}, H. and {Burke}, D.~J. and {Calderone}, G. and {Cano Rodr{\'\i}guez}, J.~L. and {Cara}, M. and {Cardoso}, J.~V.~M. and {Cheedella}, S. and {Copin}, Y. and {Corrales}, L. and {Crichton}, D. and {D'Avella}, D. and {Deil}, C. and {Depagne}, {\'E}. and {Dietrich}, J.~P. and {Donath}, A. and {Droettboom}, M. and {Earl}, N. and {Erben}, T. and {Fabbro}, S. and {Ferreira}, L.~A. and {Finethy}, T. and {Fox}, R.~T. and {Garrison}, L.~H. and {Gibbons}, S.~L.~J. and {Goldstein}, D.~A. and {Gommers}, R. and {Greco}, J.~P. and {Greenfield}, P. and {Groener}, A.~M. and {Grollier}, F. and {Hagen}, A. and {Hirst}, P. and {Homeier}, D. and {Horton}, A.~J. and {Hosseinzadeh}, G. and {Hu}, L. and {Hunkeler}, J.~S. and {Ivezi{\'c}}, {\v{Z}}. and {Jain}, A. and {Jenness}, T. and {Kanarek}, G. and {Kendrew}, S. and {Kern}, N.~S. and {Kerzendorf}, W.~E. and {Khvalko}, A. and {King}, J. and {Kirkby}, D. and {Kulkarni}, A.~M. and {Kumar}, A. and {Lee}, A. and {Lenz}, D. and {Littlefair}, S.~P. and {Ma}, Z. and {Macleod}, D.~M. and {Mastropietro}, M. and {McCully}, C. and {Montagnac}, S. and {Morris}, B.~M. and {Mueller}, M. and {Mumford}, S.~J. and {Muna}, D. and {Murphy}, N.~A. and {Nelson}, S. and {Nguyen}, G.~H. and {Ninan}, J.~P. and {N{\"o}the}, M. and {Ogaz}, S. and {Oh}, S. and {Parejko}, J.~K. and {Parley}, N. and {Pascual}, S. and {Patil}, R. and {Patil}, A.~A. and {Plunkett}, A.~L. and {Prochaska}, J.~X. and {Rastogi}, T. and {Reddy Janga}, V. and {Sabater}, J. and {Sakurikar}, P. and {Seifert}, M. and {Sherbert}, L.~E. and {Sherwood-Taylor}, H. and {Shih}, A.~Y. and {Sick}, J. and {Silbiger}, M.~T. and {Singanamalla}, S. and {Singer}, L.~P. and {Sladen}, P.~H. and {Sooley}, K.~A. and {Sornarajah}, S. and {Streicher}, O. and {Teuben}, P. and {Thomas}, S.~W. and {Tremblay}, G.~R. and {Turner}, J.~E.~H. and {Terr{\'o}n}, V. and {van Kerkwijk}, M.~H. and {de la Vega}, A. and {Watkins}, L.~L. and {Weaver}, B.~A. and {Whitmore}, J.~B. and {Woillez}, J. and {Zabalza}, V. and {Astropy Contributors}},
        title = "{The Astropy Project: Building an Open-science Project and Status of the v2.0 Core Package}",
      journal = {\aj},
         year = 2018,
        month = sep,
       volume = {156},
       number = {3},
          eid = {123},
        pages = {123},
          doi = {10.3847/1538-3881/aabc4f},
archivePrefix = {arXiv},
       eprint = {1801.02634},
 primaryClass = {astro-ph.IM},
       adsurl = {https://ui.adsabs.harvard.edu/abs/2018AJ....156..123A}
}

@ARTICLE{2013A&A...558A..33A,
       author = {{Astropy Collaboration} and {Robitaille}, Thomas P. and {Tollerud}, Erik J. and {Greenfield}, Perry and {Droettboom}, Michael and {Bray}, Erik and {Aldcroft}, Tom and {Davis}, Matt and {Ginsburg}, Adam and {Price-Whelan}, Adrian M. and {Kerzendorf}, Wolfgang E. and {Conley}, Alexander and {Crighton}, Neil and {Barbary}, Kyle and {Muna}, Demitri and {Ferguson}, Henry and {Grollier}, Fr{\'e}d{\'e}ric and {Parikh}, Madhura M. and {Nair}, Prasanth H. and {Unther}, Hans M. and {Deil}, Christoph and {Woillez}, Julien and {Conseil}, Simon and {Kramer}, Roban and {Turner}, James E.~H. and {Singer}, Leo and {Fox}, Ryan and {Weaver}, Benjamin A. and {Zabalza}, Victor and {Edwards}, Zachary I. and {Azalee Bostroem}, K. and {Burke}, D.~J. and {Casey}, Andrew R. and {Crawford}, Steven M. and {Dencheva}, Nadia and {Ely}, Justin and {Jenness}, Tim and {Labrie}, Kathleen and {Lim}, Pey Lian and {Pierfederici}, Francesco and {Pontzen}, Andrew and {Ptak}, Andy and {Refsdal}, Brian and {Servillat}, Mathieu and {Streicher}, Ole},
        title = "{Astropy: A community Python package for astronomy}",
      journal = {\aap},
         year = 2013,
        month = oct,
       volume = {558},
          eid = {A33},
        pages = {A33},
          doi = {10.1051/0004-6361/201322068},
archivePrefix = {arXiv},
       eprint = {1307.6212},
 primaryClass = {astro-ph.IM},
       adsurl = {https://ui.adsabs.harvard.edu/abs/2013A&A...558A..33A}
}

@ARTICLE{matplotlib,
       author = {{Hunter}, John D.},
        title = "{Matplotlib: A 2D Graphics Environment}",
      journal = {Computing in Science and Engineering},
         year = "2007",
        month = "May",
       volume = {9},
       number = {3},
        pages = {90-95},
          doi = {10.1109/MCSE.2007.55},
       adsurl = {https://ui.adsabs.harvard.edu/abs/2007CSE.....9...90H}
}

@ARTICLE{2007CSE.....9...90H,
       author = {{Hunter}, John D.},
        title = "{Matplotlib: A 2D Graphics Environment}",
      journal = {Computing in Science and Engineering},
         year = 2007,
        month = may,
       volume = {9},
       number = {3},
        pages = {90-95},
          doi = {10.1109/MCSE.2007.55},
       adsurl = {https://ui.adsabs.harvard.edu/abs/2007CSE.....9...90H}
}

@ARTICLE{scipy,
       author = {{Virtanen}, Pauli and {Gommers}, Ralf and {Oliphant},
         Travis E. and {Haberland}, Matt and {Reddy}, Tyler and
         {Cournapeau}, David and {Burovski}, Evgeni and {Peterson}, Pearu
         and {Weckesser}, Warren and {Bright}, Jonathan and {van der Walt},
         St{\'e}fan J.  and {Brett}, Matthew and {Wilson}, Joshua and
         {Jarrod Millman}, K.  and {Mayorov}, Nikolay and {Nelson}, Andrew
         R.~J. and {Jones}, Eric and {Kern}, Robert and {Larson}, Eric and
         {Carey}, CJ and {Polat}, {\.I}lhan and {Feng}, Yu and {Moore},
         Eric W. and {Vand erPlas}, Jake and {Laxalde}, Denis and
         {Perktold}, Josef and {Cimrman}, Robert and {Henriksen}, Ian and
         {Quintero}, E.~A. and {Harris}, Charles R and {Archibald}, Anne M.
         and {Ribeiro}, Ant{\^o}nio H. and {Pedregosa}, Fabian and
         {van Mulbregt}, Paul and {Contributors}, SciPy 1. 0},
        title = "{SciPy 1.0--Fundamental Algorithms for Scientific
                  Computing in Python}",
      journal = {arXiv e-prints},
         year = "2019",
        month = "Jul",
          eid = {arXiv:1907.10121},
        pages = {arXiv:1907.10121},
archivePrefix = {arXiv},
       eprint = {1907.10121},
 primaryClass = {cs.MS},
       adsurl = {https://ui.adsabs.harvard.edu/abs/2019arXiv190710121V}
}

@InProceedings{pandas,
  author    = { Wes McKinney },
  title     = { Data Structures for Statistical Computing in Python },
  booktitle = { Proceedings of the 9th Python in Science Conference },
  pages     = { 51 - 56 },
  year      = { 2010 },
  editor    = { St\'efan van der Walt and Jarrod Millman }
}

@ARTICLE{1972ApJ...176....1G,
   author = {{Gunn}, J.~E. and {Gott}, III, J.~R.},
    title = "{On the Infall of Matter Into Clusters of Galaxies and Some Effects on Their Evolution}",
  journal = {\apj},
     year = 1972,
    month = aug,
   volume = 176,
    pages = {1},
      doi = {10.1086/151605},
   adsurl = {http://adsabs.harvard.edu/abs/1972ApJ...176....1G}
}

@ARTICLE{1972ApJ...178..623T,
       author = {{Toomre}, Alar and {Toomre}, Juri},
        title = "{Galactic Bridges and Tails}",
      journal = {\apj},
         year = 1972,
        month = dec,
       volume = {178},
        pages = {623-666},
          doi = {10.1086/151823},
       adsurl = {https://ui.adsabs.harvard.edu/abs/1972ApJ...178..623T}
}

@article{1973AISAO...8....3K,
  title={The catalogue of isolated galaxies},
  author={Karachentseva, VE},
  journal={Astrofizicheskie Issledovaniia Izvestiya Spetsial'noj Astrofizicheskoj Observatorii},
  volume={8},
  pages={3--49},
  year={1973}
}

@INPROCEEDINGS{1977egsp.conf..401T,
       author = {{Toomre}, Alar},
        title = "{Mergers and Some Consequences}",
    booktitle = {Evolution of Galaxies and Stellar Populations},
         year = "1977",
       editor = {{Tinsley}, Beatrice M. and {Larson}, Richard B. Gehret, D. Campbell},
        month = "Jan",
        pages = {401},
       adsurl = {https://ui.adsabs.harvard.edu/abs/1977egsp.conf..401T}
}

@ARTICLE{1980ApJ...236..351D,
       author = {{Dressler}, A.},
        title = "{Galaxy morphology in rich clusters: implications for the formation and evolution of galaxies.}",
      journal = {\apj},
         year = 1980,
        month = mar,
       volume = {236},
        pages = {351-365},
          doi = {10.1086/157753},
       adsurl = {https://ui.adsabs.harvard.edu/abs/1980ApJ...236..351D}
}

@ARTICLE{1980ApJ...237..692L,
   author = {{Larson}, R.~B. and {Tinsley}, B.~M. and {Caldwell}, C.~N.},
    title = "{The evolution of disk galaxies and the origin of S0 galaxies}",
  journal = {\apj},
     year = 1980,
    month = may,
   volume = 237,
    pages = {692-707},
      doi = {10.1086/157917},
   adsurl = {http://adsabs.harvard.edu/abs/1980ApJ...237..692L}
}

@article{1980SvA....24..665K,
  title     = {An Analysis of the Isolated Galaxy Criterion},
  author    = {Karachentseva, V. E.},
  journal   = {Soviet Astronomy},
  volume    = {24},
  pages     = {665},
  year      = {1980},
  adsurl = {https://ui.adsabs.harvard.edu/abs/1980SvA....24..665K}
}

@ARTICLE{1981ApJ...246L...5B,
       author = {{Bhavsar}, S.~P.},
        title = "{On galaxy morphology in small groups}",
      journal = {\apjl},
         year = 1981,
        month = may,
       volume = {246},
        pages = {L5-L9},
          doi = {10.1086/183542},
       adsurl = {https://ui.adsabs.harvard.edu/abs/1981ApJ...246L...5B}
}

@ARTICLE{1995MNRAS.275...56N,
       author = {{Navarro}, Julio F. and {Frenk}, Carlos S. and {White}, Simon D.~M.},
        title = "{The assembly of galaxies in a hierarchically clustering universe}",
      journal = {\mnras},
         year = 1995,
        month = jul,
       volume = {275},
       number = {1},
        pages = {56-66},
          doi = {10.1093/mnras/275.1.56},
archivePrefix = {arXiv},
       eprint = {astro-ph/9408067},
 primaryClass = {astro-ph},
       adsurl = {https://ui.adsabs.harvard.edu/abs/1995MNRAS.275...56N}
}

@ARTICLE{1996Natur.379..613M,
   author = {{Moore}, B. and {Katz}, N. and {Lake}, G. and {Dressler}, A. and 
	{Oemler}, A.},
    title = "{Galaxy harassment and the evolution of clusters of galaxies}",
  journal = {\nat},
   eprint = {astro-ph/9510034},
     year = 1996,
    month = feb,
   volume = 379,
    pages = {613-616},
      doi = {10.1038/379613a0},
   adsurl = {http://adsabs.harvard.edu/abs/1996Natur.379..613M}
}

@ARTICLE{1996astro.ph.11148B,
       author = {{Bahcall}, Neta A.},
        title = "{Clusters and superclusters of galaxies}",
      journal = {arXiv e-prints},
         year = 1996,
        month = nov,
          eid = {astro-ph/9611148},
        pages = {astro-ph/9611148},
          doi = {10.48550/arXiv.astro-ph/9611148},
archivePrefix = {arXiv},
       eprint = {astro-ph/9611148},
 primaryClass = {astro-ph},
       adsurl = {https://ui.adsabs.harvard.edu/abs/1996astro.ph.11148B}
}

@ARTICLE{1998A&A...331L...1S,
   author = {{Silk}, J. and {Rees}, M.~J.},
    title = "{Quasars and galaxy formation}",
  journal = {\aap},
   eprint = {astro-ph/9801013},
     year = 1998,
    month = mar,
   volume = 331,
    pages = {L1-L4},
   adsurl = {http://adsabs.harvard.edu/abs/1998A%26A...331L...1S}
}

@ARTICLE{2001AJ....121..808C,
       author = {{Colbert}, James W. and {Mulchaey}, John S. and {Zabludoff}, Ann I.},
        title = "{The Optical and Near-Infrared Morphologies of Isolated Early-Type Galaxies}",
      journal = {\aj},
         year = 2001,
        month = feb,
       volume = {121},
       number = {2},
        pages = {808-819},
          doi = {10.1086/318758},
archivePrefix = {arXiv},
       eprint = {astro-ph/0010534},
 primaryClass = {astro-ph},
       adsurl = {https://ui.adsabs.harvard.edu/abs/2001AJ....121..808C}
}

@ARTICLE{2003ApJ...598..260P,
       author = {{Prada}, Francisco and {Vitvitska}, Mayrita and {Klypin}, Anatoly and {Holtzman}, Jon A. and {Schlegel}, David J. and {Grebel}, Eva K. and {Rix}, H. -W. and {Brinkmann}, J. and {McKay}, T.~A. and {Csabai}, I.},
        title = "{Observing the Dark Matter Density Profile of Isolated Galaxies}",
      journal = {\apj},
         year = 2003,
        month = nov,
       volume = {598},
       number = {1},
        pages = {260-271},
          doi = {10.1086/378669},
archivePrefix = {arXiv},
       eprint = {astro-ph/0301360},
 primaryClass = {astro-ph},
       adsurl = {https://ui.adsabs.harvard.edu/abs/2003ApJ...598..260P}
}

@ARTICLE{2004MNRAS.354..851R,
       author = {{Reda}, Fatma M. and {Forbes}, Duncan A. and {Beasley}, Michael A. and {O'Sullivan}, Ewan J. and {Goudfrooij}, Paul},
        title = "{The photometric properties of isolated early-type galaxies}",
      journal = {\mnras},
         year = 2004,
        month = nov,
       volume = {354},
       number = {3},
        pages = {851-869},
          doi = {10.1111/j.1365-2966.2004.08250.x},
archivePrefix = {arXiv},
       eprint = {astro-ph/0408424},
 primaryClass = {astro-ph},
       adsurl = {https://ui.adsabs.harvard.edu/abs/2004MNRAS.354..851R}
}

@ARTICLE{2005AJ....129.2562B,
       author = {{Blanton}, Michael R. and {Schlegel}, David J. and
         {Strauss}, Michael A. and {Brinkmann}, J. and {Finkbeiner}, Douglas and
         {Fukugita}, Masataka and {Gunn}, James E. and {Hogg}, David W. and
         {Ivezi{\'c}}, {\v{Z}}eljko and {Knapp}, G.~R. and {Lupton}, Robert H. and
         {Munn}, Jeffrey A. and {Schneider}, Donald P. and {Tegmark}, Max and
         {Zehavi}, Idit},
        title = "{New York University Value-Added Galaxy Catalog: A Galaxy Catalog Based on New Public Surveys}",
      journal = {\aj},
         year = "2005",
        month = "Jun",
       volume = {129},
       number = {6},
        pages = {2562-2578},
          doi = {10.1086/429803},
archivePrefix = {arXiv},
       eprint = {astro-ph/0410166},
 primaryClass = {astro-ph},
       adsurl = {https://ui.adsabs.harvard.edu/abs/2005AJ....129.2562B}
}

@ARTICLE{2005MNRAS.356.1155C,
       author = {{Croton}, Darren J. and {Farrar}, Glennys R. and {Norberg}, Peder and {Colless}, Matthew and {Peacock}, John A. and {Baldry}, I.~K. and {Baugh}, C.~M. and {Bland-Hawthorn}, J. and {Bridges}, T. and {Cannon}, R. and {Cole}, S. and {Collins}, C. and {Couch}, W. and {Dalton}, G. and {De Propris}, R. and {Driver}, S.~P. and {Efstathiou}, G. and {Ellis}, R.~S. and {Frenk}, C.~S. and {Glazebrook}, K. and {Jackson}, C. and {Lahav}, O. and {Lewis}, I. and {Lumsden}, S. and {Maddox}, S. and {Madgwick}, D. and {Peterson}, B.~A. and {Sutherland}, W. and {Taylor}, K.},
        title = "{The 2dF Galaxy Redshift Survey: luminosity functions by density environment and galaxy type}",
      journal = {\mnras},
         year = 2005,
        month = jan,
       volume = {356},
       number = {3},
        pages = {1155-1167},
          doi = {10.1111/j.1365-2966.2004.08546.x},
archivePrefix = {arXiv},
       eprint = {astro-ph/0407537},
 primaryClass = {astro-ph},
       adsurl = {https://ui.adsabs.harvard.edu/abs/2005MNRAS.356.1155C}
}

@ARTICLE{2005MNRAS.363....2K,
   author = {{Kere{\v s}}, D. and {Katz}, N. and {Weinberg}, D.~H. and {Dav{\'e}}, R.
	},
    title = "{How do galaxies get their gas?}",
  journal = {\mnras},
   eprint = {astro-ph/0407095},
     year = 2005,
    month = oct,
   volume = 363,
    pages = {2-28},
      doi = {10.1111/j.1365-2966.2005.09451.x},
   adsurl = {http://adsabs.harvard.edu/abs/2005MNRAS.363....2K}
}

@ARTICLE{2005Natur.433..604D,
   author = {{Di Matteo}, T. and {Springel}, V. and {Hernquist}, L.},
    title = "{Energy input from quasars regulates the growth and activity of black holes and their host galaxies}",
  journal = {\nat},
   eprint = {astro-ph/0502199},
     year = 2005,
    month = feb,
   volume = 433,
    pages = {604-607},
      doi = {10.1038/nature03335},
   adsurl = {http://adsabs.harvard.edu/abs/2005Natur.433..604D}
}

@ARTICLE{2006ApJS..162...38A,
       author = {{Adelman-McCarthy}, Jennifer K. and {Ag{\"u}eros}, Marcel A. and
         {Allam}, Sahar S. and {Anderson}, Kurt S.~J. and {Anderson}, Scott F. and
         {Annis}, James and {Bahcall}, Neta A. and {Baldry}, Ivan K. and
         {Barentine}, J.~C. and {Berlind}, Andreas and {Bernardi}, Mariangela and
         {Blanton}, Michael R. and {Boroski}, William N. and
         {Brewington}, Howard J. and {Brinchmann}, Jarle and {Brinkmann}, J. and
         {Brunner}, Robert J. and {Budav{\'a}ri}, Tam{\'a}s and
         {Carey}, Larry N. and {Carr}, Michael A. and {Castander}, Francisco J. and
         {Connolly}, A.~J. and {Csabai}, Istv{\'a}n and {Czarapata}, Paul C. and
         {Dalcanton}, Julianne J. and {Doi}, Mamoru and {Dong}, Feng and
         {Eisenstein}, Daniel J. and {Evans}, Michael L. and {Fan}, Xiaohui and
         {Finkbeiner}, Douglas P. and {Friedman}, Scott D. and
         {Frieman}, Joshua A. and {Fukugita}, Masataka and {Gillespie}, Bruce and
         {Glazebrook}, Karl and {Gray}, Jim and {Grebel}, Eva K. and
         {Gunn}, James E. and {Gurbani}, Vijay K. and {de Haas}, Ernst and
         {Hall}, Patrick B. and {Harris}, Frederick H. and {Harvanek}, Michael and
         {Hawley}, Suzanne L. and {Hayes}, Jeffrey and {Hendry}, John S. and
         {Hennessy}, Gregory S. and {Hindsley}, Robert B. and
         {Hirata}, Christopher M. and {Hogan}, Craig J. and {Hogg}, David W. and
         {Holmgren}, Donald J. and {Holtzman}, Jon A. and {Ichikawa}, Shin-ichi and
         {Ivezi{\'c}}, {\v{Z}}eljko and {Jester}, Sebastian and
         {Johnston}, David E. and {Jorgensen}, Anders M. and {Juri{\'c}}, Mario and
         {Kent}, Stephen M. and {Kleinman}, S.~J. and {Knapp}, G.~R. and
         {Kniazev}, Alexei Yu. and {Kron}, Richard G. and {Krzesinski}, Jurek and
         {Kuropatkin}, Nikolay and {Lamb}, Donald Q. and {Lampeitl}, Hubert and
         {Lee}, Brian C. and {Leger}, R. French and {Lin}, Huan and
         {Long}, Daniel C. and {Loveday}, Jon and {Lupton}, Robert H. and
         {Margon}, Bruce and {Mart{\'\i}nez-Delgado}, David and {Mand
        elbaum}, Rachel and {Matsubara}, Takahiko and {McGehee}, Peregrine M. and
         {McKay}, Timothy A. and {Meiksin}, Avery and {Munn}, Jeffrey A. and
         {Nakajima}, Reiko and {Nash}, Thomas and {Neilsen}, Eric H., Jr. and
         {Newberg}, Heidi Jo and {Newman}, Peter R. and {Nichol}, Robert C. and
         {Nicinski}, Tom and {Nieto-Santisteban}, Maria and {Nitta}, Atsuko and
         {O'Mullane}, William and {Okamura}, Sadanori and {Owen}, Russell and
         {Padmanabhan}, Nikhil and {Pauls}, George and {Peoples}, John, Jr. and
         {Pier}, Jeffrey R. and {Pope}, Adrian C. and {Pourbaix}, Dimitri and
         {Quinn}, Thomas R. and {Richards}, Gordon T. and
         {Richmond}, Michael W. and {Rockosi}, Constance M. and
         {Schlegel}, David J. and {Schneider}, Donald P. and
         {Schroeder}, Joshua and {Scranton}, Ryan and {Seljak}, Uro{\v{s}} and
         {Sheldon}, Erin and {Shimasaku}, Kazu and {Smith}, J. Allyn and
         {Smol{\v{c}}i{\'c}}, Vernesa and {Snedden}, Stephanie A. and
         {Stoughton}, Chris and {Strauss}, Michael A. and {SubbaRao}, Mark and
         {Szalay}, Alexander S. and {Szapudi}, Istv{\'a}n and {Szkody}, Paula and
         {Tegmark}, Max and {Thakar}, Aniruddha R. and {Tucker}, Douglas L. and
         {Uomoto}, Alan and {Vanden Berk}, Daniel E. and {Vandenberg}, Jan and
         {Vogeley}, Michael S. and {Voges}, Wolfgang and {Vogt}, Nicole P. and
         {Walkowicz}, Lucianne M. and {Weinberg}, David H. and
         {West}, Andrew A. and {White}, Simon D.~M. and {Xu}, Yongzhong and
         {Yanny}, Brian and {Yocum}, D.~R. and {York}, Donald G. and
         {Zehavi}, Idit and {Zibetti}, Stefano and {Zucker}, Daniel B.},
        title = "{The Fourth Data Release of the Sloan Digital Sky Survey}",
      journal = {\apjs},
         year = "2006",
        month = "Jan",
       volume = {162},
       number = {1},
        pages = {38-48},
          doi = {10.1086/497917},
archivePrefix = {arXiv},
       eprint = {astro-ph/0507711},
 primaryClass = {astro-ph},
       adsurl = {https://ui.adsabs.harvard.edu/abs/2006ApJS..162...38A}
}

@ARTICLE{2006MNRAS.373..469B,
       author = {{Baldry}, I.~K. and {Balogh}, M.~L. and {Bower}, R.~G. and {Glazebrook}, K. and {Nichol}, R.~C. and {Bamford}, S.~P. and {Budavari}, T.},
        title = "{Galaxy bimodality versus stellar mass and environment}",
      journal = {\mnras},
         year = 2006,
        month = dec,
       volume = {373},
       number = {2},
        pages = {469-483},
          doi = {10.1111/j.1365-2966.2006.11081.x},
archivePrefix = {arXiv},
       eprint = {astro-ph/0607648},
 primaryClass = {astro-ph},
       adsurl = {https://ui.adsabs.harvard.edu/abs/2006MNRAS.373..469B}
}

@ARTICLE{2007ApJ...658..898P,
       author = {{Park}, Changbom and {Choi}, Yun-Young and {Vogeley}, Michael S. and {Gott}, J. Richard, III and {Blanton}, Michael R. and {SDSS Collaboration}},
        title = "{Environmental Dependence of Properties of Galaxies in the Sloan Digital Sky Survey}",
      journal = {\apj},
         year = 2007,
        month = apr,
       volume = {658},
       number = {2},
        pages = {898-916},
          doi = {10.1086/511059},
archivePrefix = {arXiv},
       eprint = {astro-ph/0611610},
 primaryClass = {astro-ph},
       adsurl = {https://ui.adsabs.harvard.edu/abs/2007ApJ...658..898P}
}

@ARTICLE{2007ApJ...671..153Y,
   author = {{Yang}, X. and {Mo}, H.~J. and {van den Bosch}, F.~C. and {Pasquali}, A. and 
	{Li}, C. and {Barden}, M.},
    title = "{Galaxy Groups in the SDSS DR4. I. The Catalog and Basic Properties}",
  journal = {\apj},
archivePrefix = "arXiv",
   eprint = {0707.4640},
     year = 2007,
    month = dec,
   volume = 671,
    pages = {153-170},
      doi = {10.1086/522027},
   adsurl = {http://adsabs.harvard.edu/abs/2007ApJ...671..153Y}
}

@ARTICLE{2008ApJ...674L..13C,
       author = {{Cowan}, Nicolas B. and {Ivezi{\'c}}, {\v{Z}}eljko},
        title = "{The Environment of Galaxies at Low Redshift}",
      journal = {\apjl},
         year = 2008,
        month = feb,
       volume = {674},
       number = {1},
        pages = {L13},
          doi = {10.1086/528986},
archivePrefix = {arXiv},
       eprint = {0801.0312},
 primaryClass = {astro-ph},
       adsurl = {https://ui.adsabs.harvard.edu/abs/2008ApJ...674L..13C}
}

@ARTICLE{2008ApJS..175..356H,
   author = {{Hopkins}, P.~F. and {Hernquist}, L. and {Cox}, T.~J. and {Kere{\v s}}, D.
	},
    title = "{A Cosmological Framework for the Co-Evolution of Quasars, Supermassive Black Holes, and Elliptical Galaxies. I. Galaxy Mergers and Quasar Activity}",
  journal = {\apjs},
archivePrefix = "arXiv",
   eprint = {0706.1243},
     year = 2008,
    month = apr,
   volume = 175,
    pages = {356-389},
      doi = {10.1086/524362},
   adsurl = {http://adsabs.harvard.edu/abs/2008ApJS..175..356H}
}

@ARTICLE{2008MNRAS.384..386C,
       author = {{Cox}, T.~J. and {Jonsson}, Patrik and {Somerville}, Rachel S. and {Primack}, Joel R. and {Dekel}, Avishai},
        title = "{The effect of galaxy mass ratio on merger-driven starbursts}",
      journal = {\mnras},
         year = 2008,
        month = feb,
       volume = {384},
       number = {1},
        pages = {386-409},
          doi = {10.1111/j.1365-2966.2007.12730.x},
archivePrefix = {arXiv},
       eprint = {0709.3511},
 primaryClass = {astro-ph},
       adsurl = {https://ui.adsabs.harvard.edu/abs/2008MNRAS.384..386C}
}

@ARTICLE{2009A&A...498..407G,
       author = {{Grossi}, M. and {di Serego Alighieri}, S. and {Giovanardi}, C. and {Gavazzi}, G. and {Giovanelli}, R. and {Haynes}, M.~P. and {Kent}, B.~R. and {Pellegrini}, S. and {Stierwalt}, S. and {Trinchieri}, G.},
        title = "{The Hi content of early-type galaxies from the ALFALFA survey. II. The case of low density environments}",
      journal = {\aap},
         year = 2009,
        month = may,
       volume = {498},
       number = {2},
        pages = {407-417},
          doi = {10.1051/0004-6361/200810823},
archivePrefix = {arXiv},
       eprint = {0903.0602},
 primaryClass = {astro-ph.CO},
       adsurl = {https://ui.adsabs.harvard.edu/abs/2009A&A...498..407G}
}

@ARTICLE{2009ApJ...690.1883G,
       author = {{Gallazzi}, Anna and {Bell}, Eric F. and {Wolf}, Christian and {Gray}, Meghan E. and {Papovich}, Casey and {Barden}, Marco and {Peng}, Chien Y. and {Meisenheimer}, Klaus and {Heymans}, Catherine and {van Kampen}, Eelco and {Gilmour}, Rachel and {Balogh}, Michael and {McIntosh}, Daniel H. and {Bacon}, David and {Barazza}, Fabio D. and {B{\"o}hm}, Asmus and {Caldwell}, John A.~R. and {H{\"a}u{\ss}ler}, Boris and {Jahnke}, Knud and {Jogee}, Shardha and {Lane}, Kyle and {Robaina}, Aday R. and {Sanchez}, Sebastian F. and {Taylor}, Andy and {Wisotzki}, Lutz and {Zheng}, Xianzhong},
        title = "{Obscured Star Formation in Intermediate-Density Environments: A Spitzer Study of the Abell 901/902 Supercluster}",
      journal = {\apj},
         year = 2009,
        month = jan,
       volume = {690},
       number = {2},
        pages = {1883-1900},
          doi = {10.1088/0004-637X/690/2/1883},
archivePrefix = {arXiv},
       eprint = {0809.2042},
 primaryClass = {astro-ph},
       adsurl = {https://ui.adsabs.harvard.edu/abs/2009ApJ...690.1883G}
}

@ARTICLE{2009ApJ...699L.178N,
       author = {{Naab}, Thorsten and {Johansson}, Peter H. and {Ostriker}, Jeremiah P.},
        title = "{Minor Mergers and the Size Evolution of Elliptical Galaxies}",
      journal = {\apjl},
         year = 2009,
        month = jul,
       volume = {699},
       number = {2},
        pages = {L178-L182},
          doi = {10.1088/0004-637X/699/2/L178},
archivePrefix = {arXiv},
       eprint = {0903.1636},
 primaryClass = {astro-ph.CO},
       adsurl = {https://ui.adsabs.harvard.edu/abs/2009ApJ...699L.178N}
}

@ARTICLE{2010ApJ...721..193P,
   author = {{Peng}, Y.-j. and {Lilly}, S.~J. and {Kova{\v c}}, K. and {Bolzonella}, M. and 
	{Pozzetti}, L. and {Renzini}, A. and {Zamorani}, G. and {Ilbert}, O. and 
	{Knobel}, C. and {Iovino}, A. and {Maier}, C. and {Cucciati}, O. and 
	{Tasca}, L. and {Carollo}, C.~M. and {Silverman}, J. and {Kampczyk}, P. and 
	{de Ravel}, L. and {Sanders}, D. and {Scoville}, N. and {Contini}, T. and 
	{Mainieri}, V. and {Scodeggio}, M. and {Kneib}, J.-P. and {Le F{\`e}vre}, O. and 
	{Bardelli}, S. and {Bongiorno}, A. and {Caputi}, K. and {Coppa}, G. and 
	{de la Torre}, S. and {Franzetti}, P. and {Garilli}, B. and 
	{Lamareille}, F. and {Le Borgne}, J.-F. and {Le Brun}, V. and 
	{Mignoli}, M. and {Perez Montero}, E. and {Pello}, R. and {Ricciardelli}, E. and 
	{Tanaka}, M. and {Tresse}, L. and {Vergani}, D. and {Welikala}, N. and 
	{Zucca}, E. and {Oesch}, P. and {Abbas}, U. and {Barnes}, L. and 
	{Bordoloi}, R. and {Bottini}, D. and {Cappi}, A. and {Cassata}, P. and 
	{Cimatti}, A. and {Fumana}, M. and {Hasinger}, G. and {Koekemoer}, A. and 
	{Leauthaud}, A. and {Maccagni}, D. and {Marinoni}, C. and {McCracken}, H. and 
	{Memeo}, P. and {Meneux}, B. and {Nair}, P. and {Porciani}, C. and 
	{Presotto}, V. and {Scaramella}, R.},
    title = "{Mass and Environment as Drivers of Galaxy Evolution in SDSS and zCOSMOS and the Origin of the Schechter Function}",
  journal = {\apj},
archivePrefix = "arXiv",
   eprint = {1003.4747},
 primaryClass = "astro-ph.CO",
     year = 2010,
    month = sep,
   volume = 721,
    pages = {193-221},
      doi = {10.1088/0004-637X/721/1/193},
   adsurl = {http://adsabs.harvard.edu/abs/2010ApJ...721..193P}
}

@ARTICLE{2011MNRAS.410..417S,
       author = {{Skibba}, Ramin A. and {van den Bosch}, Frank C. and {Yang}, Xiaohu and {More}, Surhud and {Mo}, Houjun and {Fontanot}, Fabio},
        title = "{Are brightest halo galaxies central galaxies?}",
      journal = {\mnras},
         year = 2011,
        month = jan,
       volume = {410},
       number = {1},
        pages = {417-431},
          doi = {10.1111/j.1365-2966.2010.17452.x},
archivePrefix = {arXiv},
       eprint = {1001.4533},
 primaryClass = {astro-ph.CO},
       adsurl = {https://ui.adsabs.harvard.edu/abs/2011MNRAS.410..417S}
}

@ARTICLE{2011MNRAS.411..929G,
       author = {{Gr{\"u}tzbauch}, Ruth and {Conselice}, Christopher J. and {Varela}, Jes{\'u}s and {Bundy}, Kevin and {Cooper}, Michael C. and {Skibba}, Ramin and {Willmer}, Christopher N.~A.},
        title = "{How does galaxy environment matter? The relationship between galaxy environments, colour and stellar mass at 0.4 < z < 1 in the Palomar/DEEP2 survey}",
      journal = {\mnras},
         year = 2011,
        month = feb,
       volume = {411},
       number = {2},
        pages = {929-946},
          doi = {10.1111/j.1365-2966.2010.17727.x},
archivePrefix = {arXiv},
       eprint = {1009.3189},
 primaryClass = {astro-ph.CO},
       adsurl = {https://ui.adsabs.harvard.edu/abs/2011MNRAS.411..929G}
}

@ARTICLE{2011MNRAS.411.1869L,
       author = {{Li}, I.~H. and {Glazebrook}, Karl and {Gilbank}, David and {Balogh}, Michael and {Bower}, Richard and {Baldry}, Ivan and {Davies}, Greg and {Hau}, George and {McCarthy}, Pat},
        title = "{Dependence of star formation activity on stellar mass and environment from the Redshift One LDSS-3 Emission line Survey}",
      journal = {\mnras},
         year = 2011,
        month = mar,
       volume = {411},
       number = {3},
        pages = {1869-1879},
          doi = {10.1111/j.1365-2966.2010.17816.x},
archivePrefix = {arXiv},
       eprint = {1010.1447},
 primaryClass = {astro-ph.CO},
       adsurl = {https://ui.adsabs.harvard.edu/abs/2011MNRAS.411.1869L}
}

@ARTICLE{2012ARA&A..50..455F,
       author = {{Fabian}, A.~C.},
        title = "{Observational Evidence of Active Galactic Nuclei Feedback}",
      journal = {\araa},
         year = 2012,
        month = sep,
       volume = {50},
        pages = {455-489},
          doi = {10.1146/annurev-astro-081811-125521},
archivePrefix = {arXiv},
       eprint = {1204.4114},
 primaryClass = {astro-ph.CO},
       adsurl = {https://ui.adsabs.harvard.edu/abs/2012ARA&A..50..455F}
}

@ARTICLE{2012ApJ...757....4P,
       author = {{Peng}, Ying-jie and {Lilly}, Simon J. and {Renzini}, Alvio and
        {Carollo}, Marcella},
        title = "{Mass and Environment as Drivers of Galaxy Evolution. II. The Quenching
        of Satellite Galaxies as the Origin of Environmental Effects}",
      journal = {\apj},
         year = 2012,
        month = Sep,
       volume = {757},
          eid = {4},
        pages = {4},
          doi = {10.1088/0004-637X/757/1/4},
archivePrefix = {arXiv},
       eprint = {1106.2546},
 primaryClass = {astro-ph.CO},
       adsurl = {https://ui.adsabs.harvard.edu/#abs/2012ApJ...757....4P}
}

@ARTICLE{2012ApJ...757...85G,
       author = {{Geha}, M. and {Blanton}, M.~R. and {Yan}, R. and {Tinker}, J.~L.},
        title = "{A Stellar Mass Threshold for Quenching of Field Galaxies}",
      journal = {\apj},
         year = 2012,
        month = sep,
       volume = {757},
       number = {1},
          eid = {85},
        pages = {85},
          doi = {10.1088/0004-637X/757/1/85},
archivePrefix = {arXiv},
       eprint = {1206.3573},
 primaryClass = {astro-ph.CO},
       adsurl = {https://ui.adsabs.harvard.edu/abs/2012ApJ...757...85G}
}

@ARTICLE{2012MNRAS.419.2670M,
       author = {{Muldrew}, Stuart I. and {Croton}, Darren J. and {Skibba}, Ramin A. and {Pearce}, Frazer R. and {Ann}, Hong Bae and {Baldry}, Ivan K. and {Brough}, Sarah and {Choi}, Yun-Young and {Conselice}, Christopher J. and {Cowan}, Nicolas B. and {Gallazzi}, Anna and {Gray}, Meghan E. and {Gr{\"u}tzbauch}, Ruth and {Li}, I. -Hui and {Park}, Changbom and {Pilipenko}, Sergey V. and {Podgorzec}, Bret J. and {Robotham}, Aaron S.~G. and {Wilman}, David J. and {Yang}, Xiaohu and {Zhang}, Youcai and {Zibetti}, Stefano},
        title = "{Measures of galaxy environment - I. What is 'environment'?}",
      journal = {\mnras},
         year = 2012,
        month = jan,
       volume = {419},
       number = {3},
        pages = {2670-2682},
          doi = {10.1111/j.1365-2966.2011.19922.x},
archivePrefix = {arXiv},
       eprint = {1109.6328},
 primaryClass = {astro-ph.CO},
       adsurl = {https://ui.adsabs.harvard.edu/abs/2012MNRAS.419.2670M}
}

@ARTICLE{2012MNRAS.421.3522H,
       author = {{Hopkins}, Philip F. and {Quataert}, Eliot and {Murray}, Norman},
        title = "{Stellar feedback in galaxies and the origin of galaxy-scale winds}",
      journal = {\mnras},
         year = 2012,
        month = apr,
       volume = {421},
       number = {4},
        pages = {3522-3537},
          doi = {10.1111/j.1365-2966.2012.20593.x},
archivePrefix = {arXiv},
       eprint = {1110.4638},
 primaryClass = {astro-ph.CO},
       adsurl = {https://ui.adsabs.harvard.edu/abs/2012MNRAS.421.3522H}
}

@ARTICLE{2012MNRAS.424..232W,
       author = {{Wetzel}, Andrew R. and {Tinker}, Jeremy L. and {Conroy}, Charlie},
        title = "{Galaxy evolution in groups and clusters: star formation rates, red
        sequence fractions and the persistent bimodality}",
      journal = {\mnras},
         year = 2012,
        month = Jul,
       volume = {424},
        pages = {232-243},
          doi = {10.1111/j.1365-2966.2012.21188.x},
archivePrefix = {arXiv},
       eprint = {1107.5311},
 primaryClass = {astro-ph.CO},
       adsurl = {https://ui.adsabs.harvard.edu/#abs/2012MNRAS.424..232W}
}

@ARTICLE{2012MNRAS.424.2574W,
       author = {{Wang}, Wenting and {White}, Simon D.~M.},
        title = "{Satellite abundances around bright isolated galaxies}",
      journal = {\mnras},
         year = 2012,
        month = aug,
       volume = {424},
       number = {4},
        pages = {2574-2598},
          doi = {10.1111/j.1365-2966.2012.21256.x},
archivePrefix = {arXiv},
       eprint = {1203.0009},
 primaryClass = {astro-ph.CO},
       adsurl = {https://ui.adsabs.harvard.edu/abs/2012MNRAS.424.2574W}
}

@ARTICLE{2014MNRAS.442..533M,
       author = {{McIntosh}, Daniel H. and {Wagner}, Cory and {Cooper}, Andrew and
        {Bell}, Eric F. and {Kere{\v{s}}}, Du{\v{s}}an and {Bosch},
        Frank C. van den and {Gallazzi}, Anna and {Haines}, Tim and
        {Mann}, Justin and {Pasquali}, Anna and {Christian}, Allison M.},
        title = "{A new population of recently quenched elliptical galaxies in the SDSS}",
      journal = {\mnras},
         year = 2014,
        month = Jul,
       volume = {442},
        pages = {533-557},
          doi = {10.1093/mnras/stu808},
archivePrefix = {arXiv},
       eprint = {1308.0054},
 primaryClass = {astro-ph.GA},
       adsurl = {https://ui.adsabs.harvard.edu/#abs/2014MNRAS.442..533M}
}

@ARTICLE{2014MNRAS.442.1363W,
       author = {{Wang}, Wenting and {Sales}, Laura V. and {Henriques}, Bruno M.~B. and {White}, Simon D.~M.},
        title = "{Satellite abundances around bright isolated galaxies - II. Radial distribution and environmental effects}",
      journal = {\mnras},
         year = 2014,
        month = aug,
       volume = {442},
       number = {2},
        pages = {1363-1378},
          doi = {10.1093/mnras/stu988},
archivePrefix = {arXiv},
       eprint = {1403.2409},
 primaryClass = {astro-ph.CO},
       adsurl = {https://ui.adsabs.harvard.edu/abs/2014MNRAS.442.1363W}
}

@ARTICLE{2015ApJ...809..146B,
       author = {{Bradford}, Jeremy D. and {Geha}, Marla C. and {Blanton}, Michael R.},
        title = "{A Study in Blue: The Baryon Content of Isolated Low-mass Galaxies}",
      journal = {\apj},
         year = 2015,
        month = aug,
       volume = {809},
       number = {2},
          eid = {146},
        pages = {146},
          doi = {10.1088/0004-637X/809/2/146},
archivePrefix = {arXiv},
       eprint = {1505.04819},
 primaryClass = {astro-ph.GA},
       adsurl = {https://ui.adsabs.harvard.edu/abs/2015ApJ...809..146B}
}

@ARTICLE{2015Natur.521..192P,
       author = {{Peng}, Y. and {Maiolino}, R. and {Cochrane}, R.},
        title = "{Strangulation as the primary mechanism for shutting down star formation
        in galaxies}",
      journal = {\nat},
         year = 2015,
        month = May,
       volume = {521},
        pages = {192-195},
          doi = {10.1038/nature14439},
 primaryClass = {astro-ph.GA},
       adsurl = {https://ui.adsabs.harvard.edu/#abs/2015Natur.521..192P}
}

@ARTICLE{2017MNRAS.470.2982L,
       author = {{Lim}, S.~H. and {Mo}, H.~J. and {Lu}, Yi and {Wang}, Huiyuan and {Yang}, Xiaohu},
        title = "{Galaxy groups in the low-redshift Universe}",
      journal = {\mnras},
         year = 2017,
        month = sep,
       volume = {470},
       number = {3},
        pages = {2982-3005},
          doi = {10.1093/mnras/stx1462},
archivePrefix = {arXiv},
       eprint = {1706.02307},
 primaryClass = {astro-ph.GA},
       adsurl = {https://ui.adsabs.harvard.edu/abs/2017MNRAS.470.2982L}
}

@ARTICLE{2018arXiv180909427F,
       author = {{Falkendal}, Theresa and {De Breuck}, Carlos and {Lehnert}, Matthew D.
        and {Drouart}, Guillaume and {Vernet}, Jo{\"e}l and {Emonts},
        Bjorn and {Lee}, Minju and {Nesvadba}, Nicole P.~H. and
        {Seymour}, Nick and {B{\'e}thermin}, Matthieu and {Kolwa},
        Sthabile and {Gullberg}, Bitten and {Wylezalek}, Dominika},
        title = "{Massive galaxies on the road to quenching: ALMA observations of powerful
        high redshift radio galaxies}",
      journal = {ArXiv e-prints},
         year = 2018,
        month = Sep,
          eid = {arXiv:1809.09427},
        pages = {arXiv:1809.09427},
archivePrefix = {arXiv},
       eprint = {1809.09427},
 primaryClass = {astro-ph.GA},
       adsurl = {https://ui.adsabs.harvard.edu/#abs/2018arXiv180909427F}
}

@ARTICLE{2019MNRAS.487.3740W,
       author = {{Wright}, Ruby J. and {Lagos}, Claudia del P. and {Davies}, Luke J.~M. and {Power}, Chris and {Trayford}, James W. and {Wong}, O. Ivy},
        title = "{Quenching time-scales of galaxies in the EAGLE simulations}",
      journal = {\mnras},
         year = 2019,
        month = aug,
       volume = {487},
       number = {3},
        pages = {3740-3758},
          doi = {10.1093/mnras/stz1410},
archivePrefix = {arXiv},
       eprint = {1810.07335},
 primaryClass = {astro-ph.GA},
       adsurl = {https://ui.adsabs.harvard.edu/abs/2019MNRAS.487.3740W}
}

@ARTICLE{2020arXiv200712200T,
       author = {{Tinker}, Jeremy L.},
        title = "{A Self-Calibrating Halo-Based Galaxy Group Finder: Algorithm and Tests}",
      journal = {arXiv e-prints},
         year = 2020,
        month = jul,
          eid = {arXiv:2007.12200},
        pages = {arXiv:2007.12200},
          doi = {10.48550/arXiv.2007.12200},
archivePrefix = {arXiv},
       eprint = {2007.12200},
 primaryClass = {astro-ph.GA},
       adsurl = {https://ui.adsabs.harvard.edu/abs/2020arXiv200712200T}
}

@ARTICLE{2021ApJ...912...45N,
       author = {{Nevin}, R. and {Blecha}, L. and {Comerford}, J. and {Greene}, J.~E. and {Law}, D.~R. and {Stark}, D.~V. and {Westfall}, K.~B. and {Vazquez-Mata}, J.~A. and {Smethurst}, R. and {Argudo-Fern{\'a}ndez}, M. and {Brownstein}, J.~R. and {Drory}, N.},
        title = "{Accurate Identification of Galaxy Mergers with Stellar Kinematics}",
      journal = {\apj},
         year = 2021,
        month = may,
       volume = {912},
       number = {1},
          eid = {45},
        pages = {45},
          doi = {10.3847/1538-4357/abe2a9},
archivePrefix = {arXiv},
       eprint = {2102.02208},
 primaryClass = {astro-ph.GA},
       adsurl = {https://ui.adsabs.harvard.edu/abs/2021ApJ...912...45N}
}

@ARTICLE{2021ApJ...923..154T,
       author = {{Tinker}, Jeremy L.},
        title = "{A Self-Calibrating Halo-Based Group Finder: Application to SDSS}",
      journal = {\apj},
         year = 2021,
        month = dec,
       volume = {923},
       number = {2},
          eid = {154},
        pages = {154},
          doi = {10.3847/1538-4357/ac2aaa},
archivePrefix = {arXiv},
       eprint = {2010.02946},
 primaryClass = {astro-ph.CO},
       adsurl = {https://ui.adsabs.harvard.edu/abs/2021ApJ...923..154T}
}

@ARTICLE{2024ApJS..272...39W,
       author = {{Wen}, Z.~L. and {Han}, J.~L.},
        title = "{A Catalog of 1.58 Million Clusters of Galaxies Identified from the DESI Legacy Imaging Surveys}",
      journal = {\apjs},
         year = 2024,
        month = jun,
       volume = {272},
       number = {2},
          eid = {39},
        pages = {39},
          doi = {10.3847/1538-4365/ad409d},
archivePrefix = {arXiv},
       eprint = {2404.02002},
 primaryClass = {astro-ph.CO},
       adsurl = {https://ui.adsabs.harvard.edu/abs/2024ApJS..272...39W}
}

@ARTICLE{AF15,
       author = {{Argudo-Fern{\'a}ndez}, M. and {Verley}, S. and {Bergond}, G. and {Duarte Puertas}, S. and {Ramos Carmona}, E. and {Sabater}, J. and {Fern{\'a}ndez Lorenzo}, M. and {Espada}, D. and {Sulentic}, J. and {Ruiz}, J.~E. and {Leon}, S.},
        title = "{Catalogues of isolated galaxies, isolated pairs, and isolated triplets in the local Universe}",
      journal = {\aap},
         year = 2015,
        month = jun,
       volume = {578},
          eid = {A110},
        pages = {A110},
          doi = {10.1051/0004-6361/201526016},
archivePrefix = {arXiv},
       eprint = {1504.00117},
 primaryClass = {astro-ph.GA},
       adsurl = {https://ui.adsabs.harvard.edu/abs/AF15}
}

@article{bentley1975multidimensional,
  title={Multidimensional binary search trees used for associative searching},
  author={Bentley, Jon Louis},
  journal={Communications of the ACM},
  volume={18},
  number={9},
  pages={509--517},
  year={1975},
  publisher={ACM New York, NY, USA}
}
\bibliographystyle{aasjournal}



\end{document}